\documentclass[sigconf]{acmart}
\AtBeginDocument{%
  \providecommand\BibTeX{{%
    \normalfont B\kern-0.5em{\scshape i\kern-0.25em b}\kern-0.8em\TeX}}}

\copyrightyear{2023} 
\acmYear{2023} 
\setcopyright{acmlicensed}\acmConference[CHI '23]{Proceedings of the 2023 CHI Conference on Human Factors in Computing Systems}{April 23--28, 2023}{Hamburg, Germany}
\acmBooktitle{Proceedings of the 2023 CHI Conference on Human Factors in Computing Systems (CHI '23), April 23--28, 2023, Hamburg, Germany}
\acmPrice{15.00}
\acmDOI{10.1145/3544548.3580808}
\acmISBN{978-1-4503-9421-5/23/04}
\usepackage{soul,color}
\usepackage{amsmath}
\usepackage{amsthm}
\usepackage{enumitem}
\usepackage{multirow}
\usepackage{balance}

\renewcommand\hl[1]{#1}

\hypersetup{
    colorlinks=false,
    pdfborder={0 0 0},
    pdfborderstyle={/S/U/W 0},
}

\begin{document}

\title[Visual Belief Elicitation Reduces the Incidence of False Discovery]{Visual Belief Elicitation Reduces the Incidence of \\False Discovery}
%Can Belief Elicitation Reduce the Incidence of False Discovery in Visualizations?


\author{Ratanond Koonchanok}
\email{rkoonch@iu.edu}
\affiliation{%
 \institution{Indiana University-Purdue University Indianapolis}
  %\city{Indianapolis}
  %\state{Indiana}
  \country{US}
}

\author{Gauri Yatindra Tawde}
\email{gtawde@iu.edu}
\affiliation{%
 \institution{Indiana University-Purdue University Indianapolis}
  %\city{Indianapolis}
  \country{US}
}

\author{Gokul Ragunandhan Narayanasamy}
\email{gokuln@iu.edu}
\affiliation{%
 \institution{Indiana University-Purdue University Indianapolis}
  %\city{Indianapolis}
  \country{US}
}

\author{Shalmali Walimbe}
\email{swalimbe@iu.edu}
\affiliation{%
 \institution{Indiana University-Purdue University Indianapolis}
  %\city{Indianapolis}
  \country{US}
}

\author{Khairi Reda}
\email{redak@iu.edu}
\affiliation{%
 \institution{Indiana University-Purdue University Indianapolis}
  %\city{Indianapolis}
  \country{US}
}

\renewcommand{\shortauthors}{Koonchanok et al.}

\begin{abstract}
Visualization supports exploratory data analysis (EDA), but EDA frequently presents spurious charts, which can mislead people into drawing unwarranted conclusions. We investigate interventions to prevent false discovery from visualized data. We evaluate whether eliciting analyst beliefs helps guard against the over-interpretation of noisy visualizations. In two experiments, we exposed participants to both spurious and `true' scatterplots, and assessed their ability to infer data-generating models that underlie those samples. Participants who underwent prior belief elicitation made 21\% more correct inferences along with 12\% fewer false discoveries. This benefit was observed across a variety of sample characteristics, suggesting broad utility to the intervention. However, additional interventions to highlight counterevidence and sample uncertainty did not provide significant advantage. Our findings suggest that lightweight, belief-driven interactions can yield a reliable, if moderate, reduction in false discovery. This work also suggests future directions to improve visual inference and reduce bias. The data and materials for this paper are available at \url{https://osf.io/52u6v/}
\end{abstract}




\maketitle

\section{Introduction}

Interactive visualization systems are increasingly seen as essential tools in the data science ecosystem. A guiding philosophy behind the design of these tools is to facilitate analysis at the speed of sight~\cite{card1999readings}.  Accordingly, visualization systems have been designed to allow quick and almost effortless exploration of data. A key implicit metric for judging the utility of these systems is how quickly they allow users to slice and dice  data~\cite{liu2014effects}, so as to generate as many insights as possible~\cite{north2011comparison,reda2015effects}. This design philosophy is historically appropriate when one considers the role of visualization in exploratory data analysis (EDA)~\cite{tukey1977exploratory}. Yet, it is clear that people use (or are encouraged to use) visualizations for more than just canonical EDA~\cite{thomas2006visual}. Commercial visualization systems like Tableau are marketed as tools to aid people in `forecasting' and `decision-making'. There is thus tacit acknowledgment that people will or should use these systems not just to assess a dataset at hand, but also to infer something more generalizable. Statistical inference from visualized data is indeed possible~\cite{buja2009statistical}. However, one should take care to account for various sources of uncertainty, including how likely a visual pattern is to represent true effects versus accumulation of noise. Because data almost always represents a limited and potentially biased sample, it often presents spurious signals. The latter could manifest as persuasive visualizations when plotted. While some patterns (e.g., the correlation between shark attacks and tornadoes~\cite{vigen2015spurious}) are easy to dismiss as chance, other spurious visualizations might lend a convincing, if ultimately false, interpretation. 

The issue of false discovery is heightened during interactive visual analysis. As data is iteratively sliced, diced, and plotted, the chance of surfacing a spurious pattern is increased due to the so-called `multiple comparisons' problem. Systems that prioritize quick exploration could thus mislead observers into perceiving `insights' from what might be noise. In one experiment, Zgraggen et al. found that up to 60\% of insights generated through visual analysis are false~\cite{zgraggen2018investigating}. The risk of false discovery is often controlled during formal statistical modeling, but few of these methods are appropriate for an interactive analysis regime~\cite{zhao2017controlling,buja2009statistical}. Furthermore, there is limited empirical evaluation of how intuitive these techniques are, and whether human analysts can effectively leverage them to attain a reduction in false discovery. This gap takes urgency when one considers the possible role of interactive analytics systems in fueling a ``replication crises''~\cite{ioannidis2005most,munafo2017manifesto}.

An intervention that we explore in this work is inviting users to share their mental models and hypotheses with the system.  This can be done by asking the viewer to `paint' or otherwise visually specify a pattern they expect to see in a visualization before the actual data is revealed. Graphical belief elicitation has been investigated for its effect on viewer engagement~\cite{kim2017explaining,heyer2020pushing} and for promoting good analysis practices~\cite{koonchanok2021data}. In addition to these benefits, we hypothesize that visual belief elicitation is a viable intervention for neutralizing (or at least reducing) the danger of spurious visualizations. The idea is that by nudging analysts to weigh their prior knowledge during the inference process, we help guard against extreme or otherwise spurious samples. Effectively, belief elicitation could prevent the analyst from overfitting a noisy visualization. This in turn should reduce the false discovery rate (FDR). Or so we conjecture.

We conducted two crowdsourced experiments to test the above hypothesis. Participants in our experiments saw scatterplots that were sampled from known linear models and were asked to articulate the true relationship underlying those samples. We controlled sample characteristics, including size and the sample's congruence with the ground truth (i.e., whether the sample reflected the real model or a spurious pattern). We find that eliciting participant's beliefs \emph{before} displaying a  sample leads to better inference. Specifically, those who underwent prior elicitation were 21\% more likely to articulate the true model than those who just provided their updated belief \emph{after} observing the sample. We also saw a roughly 12\% reduction in the rate of false discoveries for our intervention relative to the control. In a second experiment, we investigated additional interventions, aiming to reduce the potential for confirmation bias and to better communicate a sample's predictive utility. Relative to the primary intervention, these additional encodings did not lead to any better or worse inference. Our findings suggest that belief elicitation can be a broadly useful intervention to combat the problem of spurious discovery. If integrated into general-purpose visualization systems, the interactions we tested may help reduce the incidence of false discovery in visual analytics. Our findings also suggest a need for future research into potential side effects to eliciting analyst beliefs, as well as techniques to help observers better contextualize sample robustness into their visual inference.

\section{Background \& Related Work}

\subsection{Graphical Inference and False Discovery}

Statistical inference allows for generalizing an observed (or assumed) result from a sample to a population~\cite{lowry2014concepts}. In null hypothesis significance testing (NHST), one typically tests the probability of encountering a sample that is as extreme as the observed result under a `null' hypothesis. The null typically represents a lack of an interesting effect or relationship. By design, NHST admits a percentage of results that would be incorrectly declared as significant discoveries, when in reality they are due to chance. The percentage of false discoveries admitted, referred to as $\alpha$-value, is   customarily set to 5\% in scientific publications. While the chance of false discovery will not exceed 5\% in a single test, the probability is quickly inflated as one conducts additional tests. For example, after just 10 inferences, the probability of erroneously admitting at least one discovery is equal to $1-(1-\alpha)^{10}=1-(0.95)^{10} \approx 40\%$. This issue is known as the multiple comparisons problem~\cite{althouse2016adjust} and is typically addressed in statistical modeling, for example, using Bonferroni correction or the Benjamini-Hochberg procedure~\cite{benjamini1995controlling}.

There is an analogy between NHST and visual analysis. When one inspects a visualization, they are implicitly looking for something of interest~\cite{zgraggen2018investigating}. Perhaps a non-zero correlation in a scatterplot, or a difference between two bars in a grouped bar chart. When one discovers a visualization of interest, it  means that the visualization stands out in some unexpected way. This is similar to the NHST regime where one tests a dataset to see if it supports a hypothesis of interest against a null hypothesis~\cite{buja2009statistical}. From this analogy, it follows that one should account for the likelihood of obtaining a spurious visualization that is at least as extreme as the one being observed. Buja et al. proposed the `lineup' protocol as a method to ensure a certain $\alpha$ threshold when making graphical inferences~\cite{buja2009statistical}. Their methods work by concealing a plot of the real data among a set of decoys that had been generated from a null model. A naive observer who is able to correctly identify the real plot provides statistical evidence of a difference between the true and the null generating processes. Majumder et al. validated this method with humans, finding that observers can sometimes outperform statistical inference methods~\cite{majumder2013validation}. This work suggests that, given sufficient tools, people can make reliable inferences from visualizations and even beat a statistical machine under certain conditions.

Endowing graphs with an inferential method like the lineup protocol serves to bridge the gulf between exploratory and confirmatory techniques~\cite{majumder2013validation,cook2021foundation}. A potential challenge in using the lineup method in practice lies in finding a credible null model. Real data can be complex, often embodying more structure than the assumptions of a simple null distribution. For example, permutation procedures suggested by Buja et al.~\cite{buja1987data} are insufficient at modeling real-world phenomena such as spatial-auto correlation~\cite{beecham2016map}, leading to lineups in which the `answer' is rather obvious. This could leave the analyst with a false sense of confidence. Moreover, there can be many plausible null hypotheses to test given a single dataset, making necessary multiple lineups. On the other hand, ignoring the risk of spurious visual patterns altogether can lead to a high number of false discoveries. Zgraggen et al. show that it is possible to conduct post-hoc correction for multiple comparisons after a visual analysis session~\cite{zgraggen2018investigating}. However, their method currently requires a manual review of analyst interactions and eye gaze behaviors in order to account for the implicit visual tests performed.  Zhao et al. propose a scheme in which the analyst decides where to `invest' their $\alpha$ balance as they visually test multiple hypotheses~\cite{zhao2017controlling}. Savvides et al. propose splitting data to test interesting visual patterns for significance, or alternatively limiting the set of plausible hypotheses based on the analyst's prior knowledge, thus retaining statistical power~\cite{savvides2019significance}. They subsequently propose two methods to control the FDR for between- and within-view comparisons~\cite{savvides2022visual}. Although promising, these approaches are yet to be validated by human analysts for usability and/or effectiveness. 

\subsection{Role of Prior Knowledge in Visualization Interpretation}

An alternative to the NHST framing is to consider the role of analyst beliefs in the inference process. People naturally draw upon their prior knowledge and existing models as they attempt to make sense of data~\cite{klein2007data}, with existing frames tested and adapted to explain new observations~\cite{klein2006making}. Choi et al. suggest that the same is true in visual analysis; people appear to use visualizations to test and refine their models, more often than they seek to acquire new models from data~\cite{choi2019concept}. Gelman formalizes this as a \emph{model-check} process: the viewer compares the visualization to an imaginary dataset drawn from their own model, in effect checking the goodness-of-fit of the latter~\cite{gelman2003bayesian}. Hullman and Gelman argue that model checks govern both exploratory and confirmatory analyses~\cite{hullman2021designing}. Realizing the influence of prior knowledge on visualization interpretation, designers have experimented with interactions that enable readers to externalize and `paint' their beliefs into charts~\cite{buchanan2017you}. Kim et al. found that this type of belief elicitation, when coupled with visual feedback on how well a viewer had guessed the data, can trigger better reflection and recall~\cite{kim2017explaining}. They posit that a visualization viewer performs a Bayesian inference of sort to update their belief~\cite{kim2020bayesian}.  In narrative visualization, however, eliciting beliefs and providing visual feedback did not appear to significantly impact the subjective attitudes of respondents~\cite{heyer2020pushing}. 

Techniques for eliciting beliefs are beginning to be systematically explored in the visualization and machine learning communities~\cite{mahajan2022vibe,groznik2013elicitation, daee2018user}. For example, Koonchanok et al. designed a visualization tool with belief sketching affordances~\cite{koonchanok2021data}. Participants who used that tool exhibited more normative analysis practices, such as declaring their hypotheses before peeking at the data. To streamline the process of hypothesis specification, Choi et al. developed a tool that allowed users to frame data expectations in natural language, and accordingly receive \emph{(dis)confirmatory} (i.e., model-check) visualizations~\cite{choi2019visual}. Karduni et al. tested various uncertainty representations (e.g., cones vs. lines) for how well they allowed viewers to externalize and update their models of bivariate correlation~\cite{karduni2020bayesian}. Specifying priors for visual analysis may be roughly analogous to pre-registration~\cite{nosek2018preregistration}, whereby analysts are encouraged to record their hypotheses and analysis plans before approaching the data so as to dissuade p-hacking and HARKing~\cite{yamada2018crack,p2021pre,kerr1998harking}. Pre-registration is associated with increased reporting of null results~\cite{kaplan2015likelihood}, suggesting that it may alleviate the bias in favor of positive outcomes. Formal pre-registration, however, can be difficult~\cite{nosek2019preregistration,logg2021pre} and potentially too restrictive for visual analytics where flexibility is an important consideration. Our study explores whether visual (informal) belief elicitation may provide some of the benefits of pre-registration.

This work investigates how people externalize beliefs and make visual inferences. Similar to earlier work, we employ linear, bivariate models. However, rather than measuring people's ability to perform optimal Bayesian update~\cite{karduni2020bayesian}, we study their ability to qualitatively infer data-generating models from noisy samples. We specifically focus on whether visualization viewers can evade false discovery after seeing spurious samples. In effect, we test whether visual belief elicitation is an effective intervention to prevent analysts from overfitting noisy visualizations.

\section{Research Questions \& Methods}

Earlier work suggests that elicitation of prior beliefs can positively impact the visual analytic process~\cite{koonchanok2021data,choi2019concept}. For instance, analysts appear to engage in more normative analysis practices, such as declaring hypotheses before the data is known. They also seem to adopt a more skeptical stance when given the opportunity to contrast their beliefs with data. These earlier results, however, were documented in exploratory studies. It is unclear if the behaviors above would actually lead to more reliable conclusions. Belief elicitation may indeed help address the issue of analysis reliability. Prompted to reflect on their prior knowledge, analysts may become more discerning and potentially more able to discriminate real from implausible relationships. The analyst can then consciously modulate how much they learn from a noisy visualization. In effect, the act of externalizing one's expectations about data may, in and of itself, be a good tactic to reduce overfitting and, by extension, the chance of false discovery. We seek to test this central hypothesis in this work. Specifically, we pose the following research questions: 

%\begin{itemize}[label={}]
\textbf{RQ1: } Can belief elicitation improve the accuracy of inference induced from visualizations? We prompt observers to visually register their `priors' and expectations \emph{before} revealing the data. Might this intervention lead to more accurate conclusions, particularly in the presence of noisy visualizations?

\textbf{RQ2: } Does the effect of belief elicitation depend on the reliability of the data sample at hand? A small dataset presents limited (and possibly biased) information about the ground truth. Confronted with potentially unreliable or extreme data, the analyst's prior knowledge could more strongly inform their inference, in effect reducing the impact of what could be a misleading sample. A potential side effect, however, is that belief elicitation could anchor analysts to their priors, even when provided with robust data that provides sufficient information about the ground truth. 

\textbf{RQ3: } Does highlighting a sample's predictive uncertainty or its consistency with analyst beliefs affect the reliability of insights? These two interventions could provide additional cues for analysts to decide whether to trust the data and how much to weigh their prior knowledge when making an inference.

%\end{itemize}

\begin{figure*}[t]
\center
\includegraphics[width=0.765\linewidth]{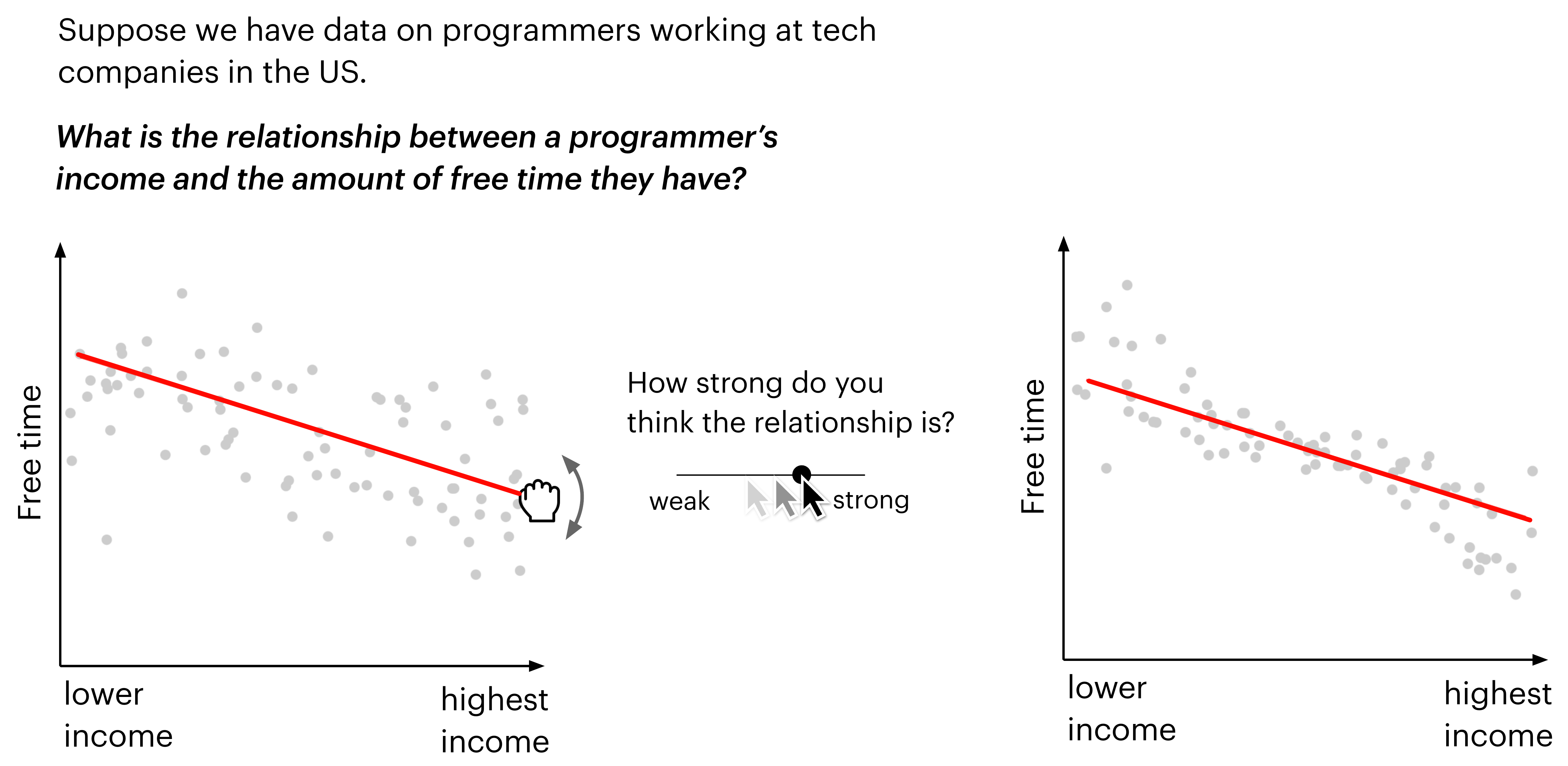}
  \caption{
  Our method for belief elicitation. A prompt sets the context (e.g., US tech workers). The participant is asked to predict the relationship between two variables (e.g., a programmer's income and the amount of free time they have). This is achieved by adjusting the slope of a trendline (red) in a scatterplot. A slider allows the participant to adjust the uncertainty in the relationship, thus controlling the spread of observations around the trendline. A sample from the model (grey points) is displayed and continuously refreshed at 5Hz, thus allowing participants to see the implications of their model. The right scatterplot reflects smaller uncertainty.}
  \label{fig:elicitation}
  \Description{
  Belief elicitation method. The figure illustrates how a participant can externalize their belief about the correlation between two variables using the interface we provide. The interface shows a question and context, such as `Suppose we have data on programmers working at tech companies in the US. What is the relationship between a programmer's income and the amount of free time they have?' The participant specifies the expected slope of the correlation by clicking and dragging a trend line. The participant then adjusts the strength of the relationship by manipulating a horizontal slider labeled `how strong do you think the relationship is'.
  }
\end{figure*}

We conducted two crowdsourced experiments to answer the above questions. Participants saw and judged datasets that had been sampled from known models. We engaged participants' prior knowledge via a graphical elicitation device. Specifically, we prompted participants to predict the parameters of the data-generating model prior to exposing them to a (potentially noisy) sample. We then captured participant inferences as open-ended responses, which we subsequently coded for accuracy against the ground truth. Rather than testing participants' ability to perform optimal Bayesian inference~\cite{kim2020bayesian,karduni2020bayesian} (a task that is difficult for many~\cite{tauber2017bayesian,cassey2016using}), our goal is to capture participants' \emph{qualitative} understanding of the data generating process. We assess whether their interpretation is consistent with the underlying statistical model. This approach acknowledges that people often hope to develop qualitative insight from visualizations~\cite{saraiya2005insight}, as opposed to arriving at an infinitesimally accurate approximation of parameters. We supplement with a post-hoc analysis of participants' posterior beliefs to understand their belief-updating process.

Another key feature of our study is controlling the rate of false-positive and false-negative datasets (i.e., samples that \emph{wrongfully} suggest the presence or lack of a relationship). This rate was controlled on a per-participant basis and kept consistent with the model likelihood. We limit our study to linear models, which dictate a relationship (or lack thereof) between two quantitative variables. We employ scatterplots for visualizing samples and for belief elicitation. We first discuss the experimental apparatus. We then describe how we synthesized ground-truth models, and how we generated true and spurious visualizations from those models.

\subsection{Model and Prior Elicitation}
\label{sec:belief_elicitation}

To elicit beliefs about linear relationships, we used a graphical device that allows observers to specify the direction and strength of the relationship. We first display a prompt question about the relationship between two variables, $x$ and $y$, and ask participants to adjust the slope of a trendline in a scatterplot to indicate the expected relationship. Additionally, participants adjust the expected uncertainty in the relationship using a slider (see Figure~\ref{fig:elicitation} for an illustration). In effect, a participant visually supplies two parameters ($\mu$ and $\sigma$) for the following linear model:

\begin{equation*}
\begin{aligned}
    y_i = \beta_0 + \beta  x_i + \epsilon_i\\
    \beta \sim \mathcal{N}(\mu, \sigma^2)\\
    \beta_0 \sim \mathcal{N}(0, \sigma_b^2) \\
    \epsilon_i \sim \mathcal{N}(0, {\sigma_e^2})
\end{aligned}
\end{equation*}

Where $\mu \in (-1, 1)$ is the slope of the relationship as specified by the trendline, and $\sigma \in (0, 1)$ is the uncertainty in the slope as specified by a slider. $\beta_0$ specifies an intercept for the regression line, centered around 0 with a fixed standard deviation of $\sigma_b=0.1$ for all stimuli. $\epsilon_i$ is an additional residual term with a standard deviation fixed to $\sigma_e=0.5$. To help participants grasp the implications of their belief, we update and sample the participant's model throughout the interaction. We display the sample in the scatterplot and continuously at 5Hz, showing a new set of points every 200 milliseconds. This animated display, which amounts to a hypothetical outcome plot (HOP), allows participants to directly see expected observations as predicted from their belief.

\begin{table*}[t]
    \centering
    \caption{Examples from the 16 prompt questions used as stimuli in our experiments. For each question, we collected a mean crowdsourced belief ($\mu_{crowd}$) and accordingly set the ground-truth model slope ($\mu$) to dictate positive, negative, or no bivariate relationship between the prompt variables. }
    \label{tab:example_questions}
    
    \begin{tabular}{c|l|c|c}
    
    \textbf{Correlation} & \textbf{Question}  & \textbf{Crowd wisdom}  & \textbf{Ground truth} \\
    %[0.5ex]
    \hline\hline

     & What is the relationship between the number of&  $\mu_{crowd}=0.40$ &   \\%[-1ex]
     
     & celebrity actors in a movie and the movie's rating? &   $\sigma_{crowd}=0.30$ & $\mu=0.5$ \\%[1ex]
     
    \cline{2-3}

     \raisebox{1ex}{Positive}& What is the relationship a singer's social media &  $\mu_{crowd}=0.50$ & $\sigma=0.29$   \\%[-1ex]
     
     & followers and their music record sales? &   $\sigma_{crowd}=0.25$ &  \\%[1ex]
          \hline

         & What is the relationship between phone screen time & $\mu_{crowd}=-0.28$ &  \\%[-1ex] 
    
     & and amount of sleep each night?  &   $\sigma_{crowd}=0.25$ & $\mu=-0.5$  \\%[1ex]
     
     \cline{2-3}
     
    \raisebox{1ex}{Negative}& What is the relationship between the number &  $\mu_{crowd}=-0.34$ & $\sigma=0.29$ \\%[-1ex]
    
     & of children in a family and the family's savings? &   $\sigma_{crowd}=0.28$ &  \\%[1ex]

     \hline
      & What is the relationship between income and &  $\mu_{crowd}=-0.08$ &   \\%[-1ex]
     
     & the amount of free time a programmer has? &   $\sigma_{crowd}=0.31$ & $\mu=0 $ \\%[1ex]
     
    \cline{2-3}

     \raisebox{1ex}{No relationship}& What is the relationship between salary &  $\mu_{crowd}=-0.06$ & $\sigma=0.29$   \\%[-1ex]
     
     & and debt for an individual in the US? &   $\sigma_{crowd}=0.29$ &  \\%[1ex]
    \end{tabular}

\end{table*}

\begin{figure*}[]
\center
\includegraphics[width=.8\linewidth]{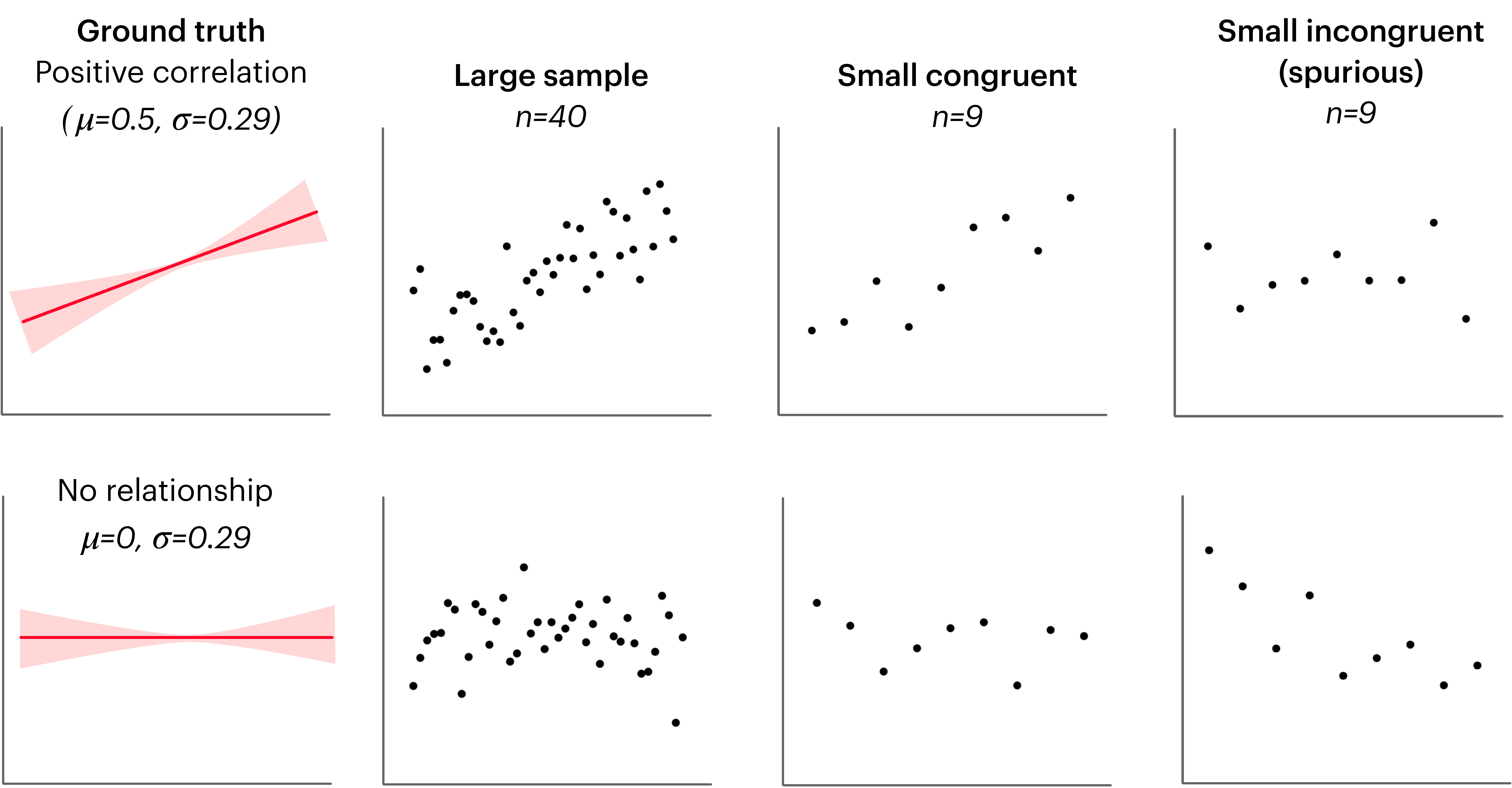}
  \caption{Examples of large (always congruent), small-congruent, and small-incongruent (i.e., spurious) samples.}
  \label{fig:samples}
  \Description{This figure shows different sample configurations as seen in our experiments. The different sample types are shown using scatterplots. The scatterplots are arranged in a 2 x 3 grid. The top row of scatterplots represents samples where the ground truth is indicating a positive correlation between two variables. A large sample (n=40) shows a scatterplot with a clear positive relationship, consistent with the ground truth. A small congruent sample (n=9) shows a scatterplot also with a positive relationship but with fewer data points. A small incongruent (i.e., spurious) sample results in a scatterplot inconsistent with the ground truth, showing no relationship between the variables (when in fact the relationship is positive).

  The second row shows 3 scatterplots for a ground truth of no correlation. The large-sample scatterplot shows no relationship, consistent with the ground truth. A small-congruent scatterplot also shows no relationship for a fewer number of points. Lastly, a small-incongruent (spurious) shows a spurious scatterplot suggesting a negative relationship, when in fact the ground truth indicates no correlation.}
  
\end{figure*}

\subsection{Seeding Ground Truth Models}
\label{sec:questions}

To test if belief elicitation helps participants make true discoveries, we devised questions to seed ground-truth models. These known models enable us to code participants' qualitative inferences (specifically, the implied bivariate relationship) for correctness. We started by formulating an initial set of 40 prompt questions, each concerning the relationship between two quantitative variables. The questions were designed to probe common knowledge: Half featured variables that were expected to show no relationship, while the other half were expected to exhibit either positive or negative correlation. Table~\ref{tab:example_questions} illustrates example questions. To empirically anchor the responses to those questions in common beliefs, we recruited 40 crowdworkers from Amazon Mechanical Turk. Each worker was tasked with providing their belief on all 40 questions using the elicitation device illustrated in Figure~\ref{fig:elicitation}. In effect, every worker provided two model parameters ($\mu$ and $\sigma$) in response to each question. We averaged the parameters across all workers (separately for each question), thus yielding a crowd wisdom response for every prompt. We subsequently used the mean slope to select a subset of 16 questions from the initial 40 to be used as stimuli in our experiments. Of those 16 questions, 8 questions demonstrated a crowd belief of no relationship between the prompt variables, 4 suggested a positive relationship, and 4 a negative relationship. In other words, half the prompts dictated a `null' ground truth while the other half specified a correlation (either positive or negative). This prompt selection was based on the crowd wisdom. Specifically, we considered questions with an average crowd slope of $\mu_{crowd}>0.26$ to reflect a wisdom of positive correlation. Accordingly, we set $\mu$ in the ground truth model for those questions to 0.5. Conversely, we considered a mean $\mu_{crowd}<-0.26$ to indicate a negative relationship and accordingly set the corresponding ground-truth model to $\mu=-0.5$. We considered questions with an average slope of $-0.12<\mu_{crowd}<0.12$ to indicate a lack of expected relationship between the two variables (i.e., a null model), setting the ground-truth slope to zero. We found that, across all questions, workers ascribed very similar uncertainty levels to their belief (${\sigma}_{crowd}$). Therefore, for all ground truth models, we set $\sigma$ to 0.29, the observed mean slope uncertainty.

\begin{table*}[]
    \caption{Characteristics of spurious (i.e., incongruent) samples seen in our simulations for the small and large sample sizes.}
    \label{tab:simulation}

\begin{tabular}{c|cc|cc}
\multicolumn{1}{c|}{\multirow{2}{*}{\textbf{\begin{tabular}[c]{@{}c@{}}Sample \\ size ($n$)\end{tabular}}}} & \multicolumn{2}{c|}{\textbf{Percent spurious samples}} & \multicolumn{2}{c}{\textbf{Mean spurious slope difference ($\Delta$)}} \\ 
\multicolumn{1}{c|}{}                                                                                 & \multicolumn{1}{@{\hspace{6mm}}c@{\hspace{6mm}}|}{$\mu=0$}        & $\mu=0.5$         & \multicolumn{1}{@{\hspace{10mm}}c@{\hspace{10mm}}|}{$\mu=0$}        & $\mu=0.5$     \\ \hline\hline
9                                                                                                      & \multicolumn{1}{c|}{37.3\%}          & 37.2\%           & \multicolumn{1}{c|}{0.273}    & 0.274    \\
40                                                                                                     & \multicolumn{1}{c|}{7.1\%}             & 5.2\%            & \multicolumn{1}{c|}{0.215}     & 0.206   
\end{tabular}
\end{table*}

\subsection{Controlling the Rate of Spurious Visualizations}
\label{sec:spurious_samples}

Participants in our study saw scatterplots sampled from the above ground-truth models and were tasked with inferring model characteristics.  We controlled two aspects of the sampling process: the sample \emph{size}, which represented the number of observations (i.e., data points) in the visualization, and the \emph{congruence} of the sample with the underlying ground truth. A sample that appears to show positive correlation is congruent with a $\mu=0.5$ model. Conversely, that same sample would be \emph{incongruent} with a model that has $\mu=0$ (i.e., no correlation in the ground truth). By the law of large numbers, incongruent samples are far more likely to arise when the sample is small. We simulate this phenomenon by including three sample types as a factor: \emph{small-incongruent}, \emph{small-congruent}, and \emph{large} (always congruent) samples. Figure~\ref{fig:samples} illustrates examples of each, as seen in our experiments. 

%We might expect observers to make correct inferences when presented with a large sample that is compatible with the model. However, the two small-sample configurations (congruent and incongruent) present a more challenging situation that could invite false discovery. By exposing participants to these samples, we can assess how useful belief elicitation might be in different sample configurations. 

\begin{figure}[t]
\center
\includegraphics[width=1\linewidth]{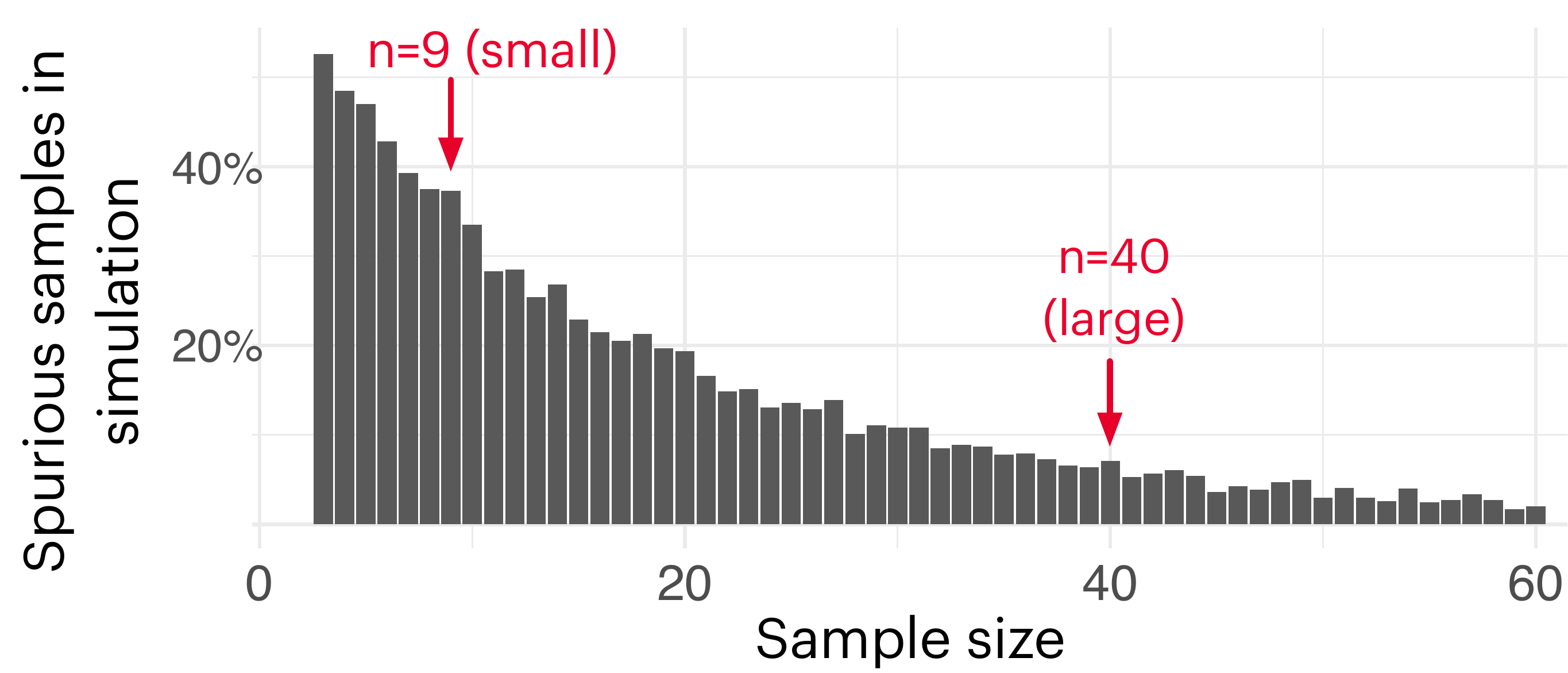}
  \caption{
  Rate of spurious samples (for $\mu=0$ models) in our simulation as a function of sample size. Arrows mark the rate for $n=9$ (`small') and $n=40$ (`large') sample sizes.}
  \label{fig:simulation}
  \Description{Rate of spurious samples. The figure shows the distribution chart of spurious samples in the simulation for the no-relationship model. The y-axis is the percentage of spurious samples ranging from 0\% to 55\% and the x-axis is the sample size ranging from 0 to 60. The chart is annotated with arrows at the selected sample size (n = 9 for small and n = 40 for large).
  }
\end{figure}

We sought to control the rate of incongruent samples such that all participants would encounter the same number of spurious visualizations. To keep the latter consistent with the likelihood as per the ground truth, we conducted simulations measuring the rate of  incongruent samples arising as a function of sample size. First, we took a model with $\mu=0$ and $\sigma=0.29$ (i.e., a \emph{no-relationship}, null model). We drew $n$-point random samples from the model, systematically varying $n$ between 3 and 60 and generating 1000 samples for each $n$ size. We fitted each sample to a linear model (of the formulation described in \S\ref{sec:belief_elicitation}) and compared the fitted slope ($\mu_{fit}$) to the ground truth $\mu$ using the absolute difference $\Delta=|\mu - \mu_{fit}|$. We considered a cutoff of $\Delta \geq 0.175$ to delineate `incongruent' samples. That is, a sample would be considered incongruent (or spurious) if the slope of its linear fit is at least 0.175 away from the real data-generating model. We chose this cutoff to yield plausible spurious visualizations while also ensuring a sufficient chance for those samples to arise. Figure~\ref{fig:simulation} illustrates the percentage of such samples in the simulation as a function of $n$. We then conducted a similar simulation for a $\mu=0.5$ ground truth (i.e., positive correlation), using the same spurious-sample cutoff of $\Delta \geq 0.175$. Based on simulation results, we selected two sample sizes: a `small' sample size of $n=9$, which yields a spurious-sample rate of 37.3\% for null models and 37.2\% for positive (and negative) correlation ground truths. We adopted $n=40$ as a `large' size, for a spurious-sample rate of 7.1\% and 5.2\%, respectively. Table~\ref{tab:simulation} provides a breakdown of those rates. We use (approximately) the same rates in our experiments and control that rate on a per-subject basis. 

In addition to providing a sufficient number of incongruent samples, the parameters above ensure similar spurious-sample characteristics in terms of how far those samples are from the ground truth, for both the null and correlation models (see Figure~\ref{fig:samples} for a visual illustration). For example, at $n=9$, the difference between the fitted slope for spurious samples and the ground truth (i.e., average $\Delta$) was 0.273 in the null vs. 0.274 in the correlation models (see Table~\ref{tab:simulation}).

\begin{figure*}[t]
\center
\includegraphics[width=1\linewidth]{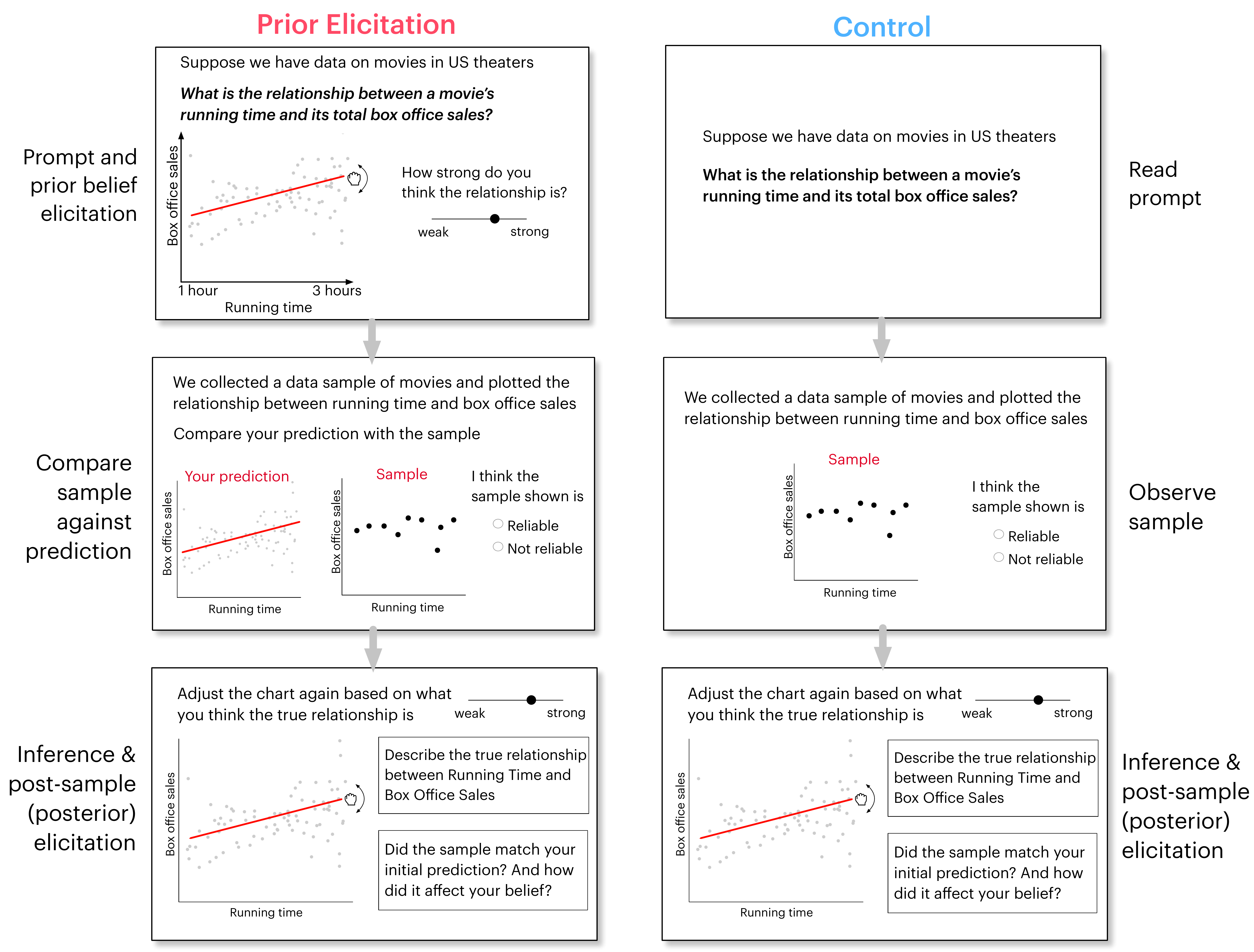}
  \caption{Illustration of the three steps that make up a trial. In the \textbf{Prior Elicitation} condition, participants first read the prompt question and specified their prior belief using a graphical elicitation device. In a second slide, they saw a sample from the ground truth visualized side-by-side with their belief. They provided their impression on whether they thought the sample was reliable. In the third and last step, the participant was asked to specify their belief once again, and respond to two open-ended prompts: to describe the true relationship and to comment on how the sample might have affected their belief. The \textbf{Control} condition followed a similar sequence with the exception that we did not elicit  beliefs on the first slide, and therefore only showed a sample on the second slide.}
  \label{fig:stimulus}
  \Description{
  Illustration of trial steps for experiment 1. The left column is the steps for the Prior Elicitation group. The first panel consists of an example question about the relationship between a movie’s running time and its total box office sales as well as the belief elicitation interface, where the participant could adjust the slope by rotating the trend line and could adjust the spread of the data points through a slider. The second panel shows the sample and prediction side by side for the participant to compare. The participant would also use a radio button to select whether they think the sample is reliable or not. The final panel shows the belief elicitation interface again for post-sample elicitation. In addition, there are two text boxes for open-ended questions. The upper box asks the participant to describe the relationship between variables. The lower box asks the participants to describe how the sample affects the relationship conclusion. The right column is the steps for the Control group. The first panel consists of the question alone. The second panel contains only the sample for the participant to observe. The third panel is identical to the one for the Prior Elicitation group.
  }
\end{figure*}

\section{Experiment I}

In this experiment, we evaluate the impact of prior belief elicitation on inferences. We present participants with a question about the relationship between two variables, and elicit their belief about the nature of that relationship (using the graphical device described in \S\ref{sec:belief_elicitation}). We then expose participants to a sample from the ground-truth model, and present that sample side-by-side with their beliefs for analysis. Following exposure to the sample, we prompt participants to make an inference about the `true' relationship, and ask them to adjust their prediction a second time. We compare this setup to a control condition that does not include  elicitation \emph{prior} to sample exposure. We score participants' inferences by manually checking for consistency with the ground truth. Our primary metric is not whether participants can do  Bayesian update, but whether their qualitative insight (captured in natural language) is compatible with the underlying data-generating model. To that end, we compare the rate of false and correct inferences in both conditions.

\subsection{Hypotheses}

We developed three hypotheses:

%\begin{itemize}[label={}]
\textbf{H1 ---} Eliciting prior beliefs will improve inference accuracy. Specifically, we expect participants in the Prior Elicitation group to have a higher number of correct inferences compared to those in Control. The argument behind this hypothesis is that eliciting participants' beliefs will nudge them to incorporate their prior beliefs in the inference, hence guarding against extreme, spurious samples. We would thus expect the intervention group to be more successful at discriminating true from spurious visualizations. 

\textbf{H2 ---} The difference in accuracy between the two groups (Prior Elicitation vs. Control) will be more pronounced when the sample is \emph{small and incongruent} with the ground truth. The latter presents a potentially higher likelihood of making an incorrect inference. We anticipate those whose beliefs are elicited prior to seeing such samples to draw on their priors, which serves to moderate the influence of misleading visualizations. 

\textbf{H3 ---} Participants who externalize their prior beliefs will be, on average, less trusting of samples relative to the Control. In particular, we expect small samples to be flagged as non-reliable more frequently in the former.

%\end{itemize}

\subsection{Participants}

We recruited 80 participants (41 male and 39 female) from Amazon Mechanical Turk. Participants had a mean age of 34.7 years. We recruited  workers who are US residents with a minimum task-approval rate of 98\%. Participants received a \$5 compensation upon completing the experiment. Based on a pilot, we estimated the experiment to take 40 minutes on average. The study was approved by Indiana University's institutional review board.

\subsection{Apparatus and Experimental Design}

The experiment was a between-subject design. Half the participants (40 individuals) were randomly assigned to the Prior Elicitation condition. The other half were assigned to the Control. Participants completed a total of 16 trials corresponding to the prompt questions developed in \S\ref{sec:questions}. Specifically, a trial consisted of one question about the relationship between two variables (e.g., ``What is the relationship between a movie's running time and its total box office sales?''). Participants in the \textbf{Prior Elicitation} condition were first asked to visually externalize their belief in response to the question. They did so by setting the slope of the trendline and adjusting the uncertainty slider (see Figure~\ref{fig:elicitation}). Next, participants were presented with a scatterplot containing a sample from the ground truth. The sample was displayed side-by-side with the belief model in one slide. The slide prompted the participant to compare the two and indicate whether they thought the sample was ``reliable'' or not. Lastly, after exposure to the sample, participants were asked to re-specify their belief graphically about the relationship, after having been exposed to a sample. Additionally, they were asked to respond to two open-ended prompts: one to report on what they \emph{inferred} about the ``true relationship'' and a second to describe how the sample affected their ``prediction''. The sequence is illustrated in Figure~\ref{fig:stimulus}. 

The \textbf{Control} condition consisted of a similar sequence although without \emph{prior} elicitation. However, and consistent with the intervention, we elicited participants' beliefs in Control \emph{after} exposure to the sample. This final elicitation step meant that the two conditions were largely comparable in their interactions, with an additional elicitation of priors in the intervention group. Post-sample (i.e., posterior) elicitation also gave participants in the Control an equal opportunity to reflect when articulating their inference, though without the benefit of having predicted a priori. In both conditions, we manipulated the size and congruence of the visualized  samples relative to the ground truth. Two of the 16 trials displayed \emph{large} samples ($n=40$ data points), with the remaining 14 showing small samples ($n=9$). Of the small samples, 40\% (6 trials) were selected to be incongruent with the underlying ground truth. This rate for spurious visualizations was approximately consistent with the model likelihood after rounding for whole numbers (see Table~\ref{tab:simulation}). The remaining small samples (8 trials) were generated to be congruent with the ground truth. We purposefully overrepresented small samples in this experiment as we sought to simulate situations where false discovery is more likely. This in turn allows us to evaluate the impact of the intervention with higher statistical power.

\subsection{Procedure}

Participants first saw a tutorial explaining the task and providing an overview of the interface. They were informed that their goal was to ``predict and then report on the \textbf{true relationship} between quantitative attributes'' (emphasis in the original prompt). Participants were instructed on how to use the belief elicitation device using a short animation. We also informed participants that the data samples they would be viewing may be ``noisy (especially when containing a few data points),'' and that they would need to think about how reliable a sample might be. 

After the tutorial, participants completed the analyzed trials. The order of trials was randomized, with the exception that the two large-sample trials were always displayed as the first and eighth stimuli (i.e., mid-experiment). We also randomly selected where (i.e., with which question) a participant would see the different sample configurations. The one constraint to this randomization is that the  small-incongruent samples would be equally represented under the null and correlation ground-truth models. Participants finished the experiment by answering a brief demographic survey. To ensure participant engagement, we read all responses and excluded from the analysis those who provided irrelevant or incomprehensible responses on more than 25\% of the trials. A response was deemed irrelevant if it did not reference the variables in question. We recruited additional participants to replace those who were excluded until we reached our intended sample size of 80 individuals.

\subsection{Coding Inference Accuracy}
\label{sec:coding}

We manually coded the correctness of inferences. Recall that, for each trial, participants were asked to describe what they thought was the ``true relationship'' between the two variables depicted. We did not provide participants with a specific template to follow. Instead, we let them express their inference as an open-ended text response. To score the accuracy of those responses, one coder (the first author) coded the type of relationship implied by the participant. Specifically, we coded the latter as indicating one of three types of relationships: positive relationship, negative relationship, or no correlation. For instance, in a question about the relationship between caffeine consumption and height, responses indicating that ``there is a negative correlation between height and caffeine,'' or that ``people who consumed a lot more caffeine tended to be shorter'' were coded as inference of a negative correlation. On other other hand, participants who concluded that ``caffeine consumption does not affect the height'' were coded as a no-relationship inference. We also coded responses that indirectly implied a relationship accordingly (e.g., ``social media is key for sales promotion'' as a positive correlation). When the response implied a relationship but did not specify a direction, we referred to the elicited, post-sample (i.e., posterior) slope to determine the sign of the relationship. Lastly, we evaluated inferences of `slight', `mild', or `very weak' relationships conservatively, coding them as  \emph{no-correlation}. For example, ``debt and salary are slightly connected'' was recorded as implying a no-correlation inference. This choice was meant to give participants the benefit of the doubt, as there were more null ground-truth models in the experiment than negative or positive correlations. We measured coding reliability by having a second coder independently code approximately 5\% of the responses. We computed Cohen's kappa to assess inter-coder agreement. The resulting kappa coefficient was 0.896, indicating strong agreement between the coders~\cite{mchugh2012interrater}. To further reduce the potential for bias during the coding process, both coders were blinded to the experimental condition.

\begin{figure}[t]
\center
\includegraphics[width=1\linewidth]{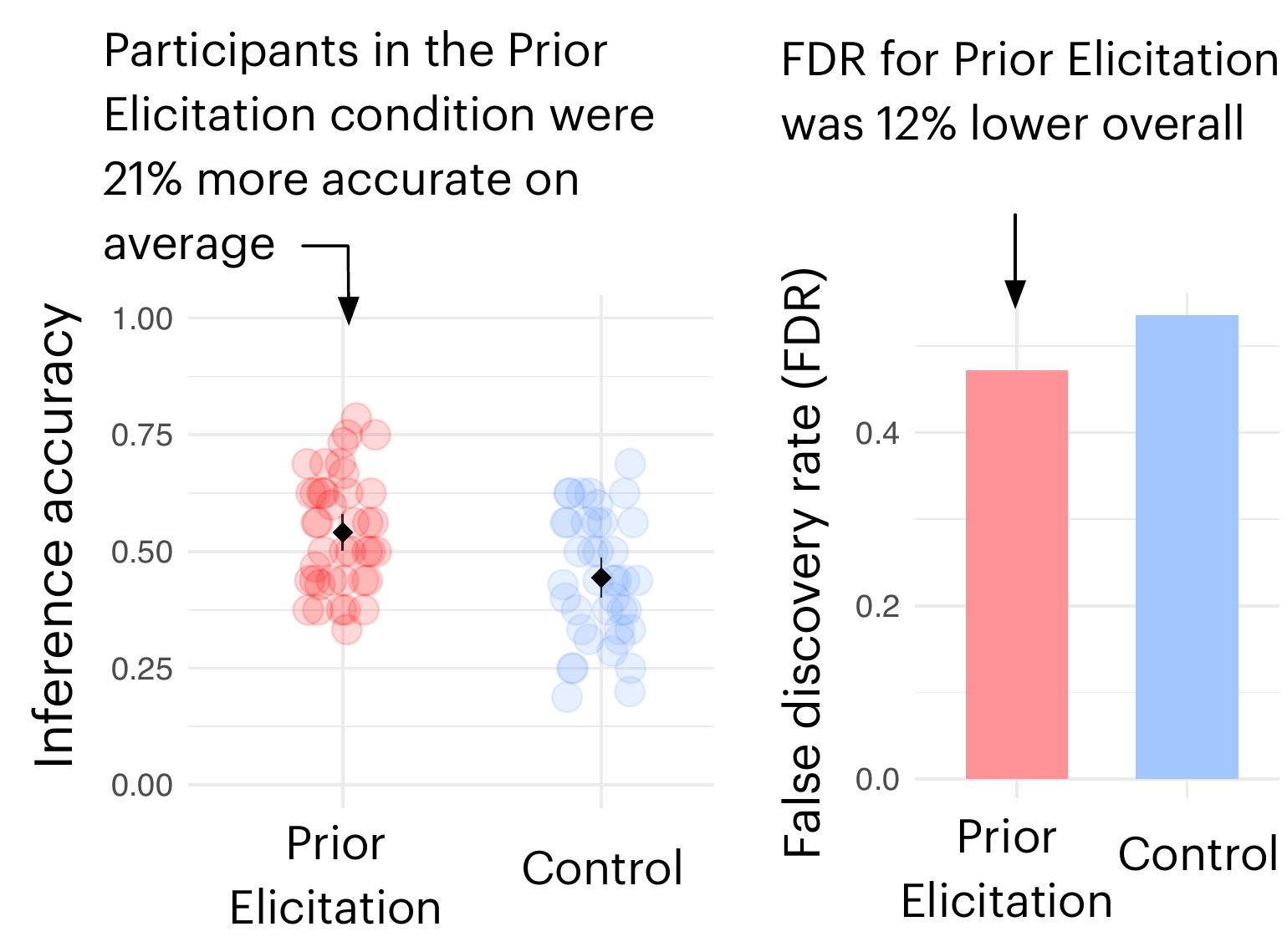}
  \caption{
  \textbf{Left:} Proportion of inferences coded  `correct' by experimental condition. Dots represent the accuracy for individual participants. Diamonds depict mean condition accuracy (with 95\% confidence intervals). \textbf{Right: } The overall false discovery rate (FDR) by condition.} 
  \label{fig:exp1_accuracy}
  \Description{
  Accuracy and false discovery rate for experiment 1. The left figure shows a vertical scatterplot comparing the inference accuracy (on the y-axis ranging from 0.00 to 1.00) for the Prior Elicitation and Control groups. Dots represent the accuracy for individual participants and diamonds depict mean condition accuracy (with 95\% confidence intervals). The right figure shows a bar chart comparing the false discovery rate (on the y-axis) for the two groups. Both figures are also annotated with a key finding at the top. The left figure is annotated "Participants in the Prior Elicitation condition were 21\% more accurate on average" and the right figure is annotated "FDR for Prior Elicitation was 12\% lower overall". 
  }
\end{figure}

Once coded, the implied direction of the relationship was compared against the ground-truth slope ($\mu$). Recall that there were three possible values for $\mu$ in our models: $\mu$=0.5 for a model of positive correlation, $-0.5$ for negative, and $0$ for no relationship. If the coded response matched the ground truth, the inference was deemed \emph{correct}. Otherwise, we deemed the inference  \emph{incorrect}. This method of scoring visualization insights against the data-generating models is similar in spirit to Zgraggen et al.~\cite{zgraggen2018investigating}. A limitation to our coding scheme, however, is that it assumes a dichotomous interpretation of inferences (i.e., `correct' vs. `incorrect'), as opposed to capturing effect size estimates~\cite{cumming2014new}. We decided to use a binary metric based on pilot data, which suggested that the majority of inferences would categorically imply the presence of a (negative or positive) relationship or the lack thereof.

\subsection{Results}

Participants completed the experiment in 53.4 minutes on average. There was no meaningful difference in completion time between the two conditions (53.7 minutes with the intervention vs. 53.1 in Control, $t(78)=.17, p=.86$). Participants collectively provided 1,280 inferences in total (half obtained under Prior Elicitation and half under Control). We excluded 28 responses ($\sim$2\% of total) that we were unable to code because they were nonsensical or did not respond to the prompt. We first analyze the correctness of inferences and then assess participants' trust in the samples shown.

\subsubsection{\underline{Inference Accuracy and False Discovery Rate}}
\label{sec:exp1_accuracy}

Using the coded inference accuracy, we fit the results to a logistic regression model. The model predicts the likelihood of a correct inference based on three fixed effects: the experimental condition (Prior Elicitation or Control), the sample type (large, small-congruent, or small-incongruent), and the ground-truth model (positive, negative, or no correlation). We included interaction terms between the experimental condition and sample type as well as between the condition and ground-truth model type. Additionally, we included two random intercepts to account for individual differences between participants as well as differences due to questions (recall that the experiment comprised 16 unique questions probing various topics). We test for significant effects using a likelihood-ratio test (relative to a reduced model) and report the associated $\chi^2$ statistic. We also report $Z$ and p-values for pairwise post-hoc tests, adjusting for multiple comparisons using Tukey's method. As an estimate of effect size, we report odds ratios and give the corresponding 95\% confidence intervals.

We found a significant main effect of the experimental condition ($\chi^2(4)=13.1, p<.05$). The odds of correct inference were 2.01 times higher (95\% CI: 1.38---2.93) for participants who underwent belief elicitation \emph{before} observing a sample ($Z=3.63, p<.001$). The intervention thus led to a higher likelihood of correctly classifying the ground truth (54\% chance, CI: 50.1---57.9\% versus 44.5\%, CI: 40.6---48.4\% in Control). The advantage amounted to 21.3\% better inference for the intervention. In addition to omnibus accuracy, we compared the \emph{false discovery rate} (FRD) in the two conditions. Recall that half the stimuli were based on a ground truth of no-relationship between the variables while the other half was grounded in a correlation (either positive or negative). FDR is the proportion of `discovered' correlations that are unfounded in the true model. The FDR for the Prior Elicitation group was 47.2\% compared to 53.5\% in Control, which amounts to an 11.7\% reduction in false discovery. Figure~\ref{fig:exp1_accuracy} illustrates the inference accuracy and FDR rates for the two conditions.

\begin{figure*}[t]
\center
\includegraphics[width=.85\linewidth]{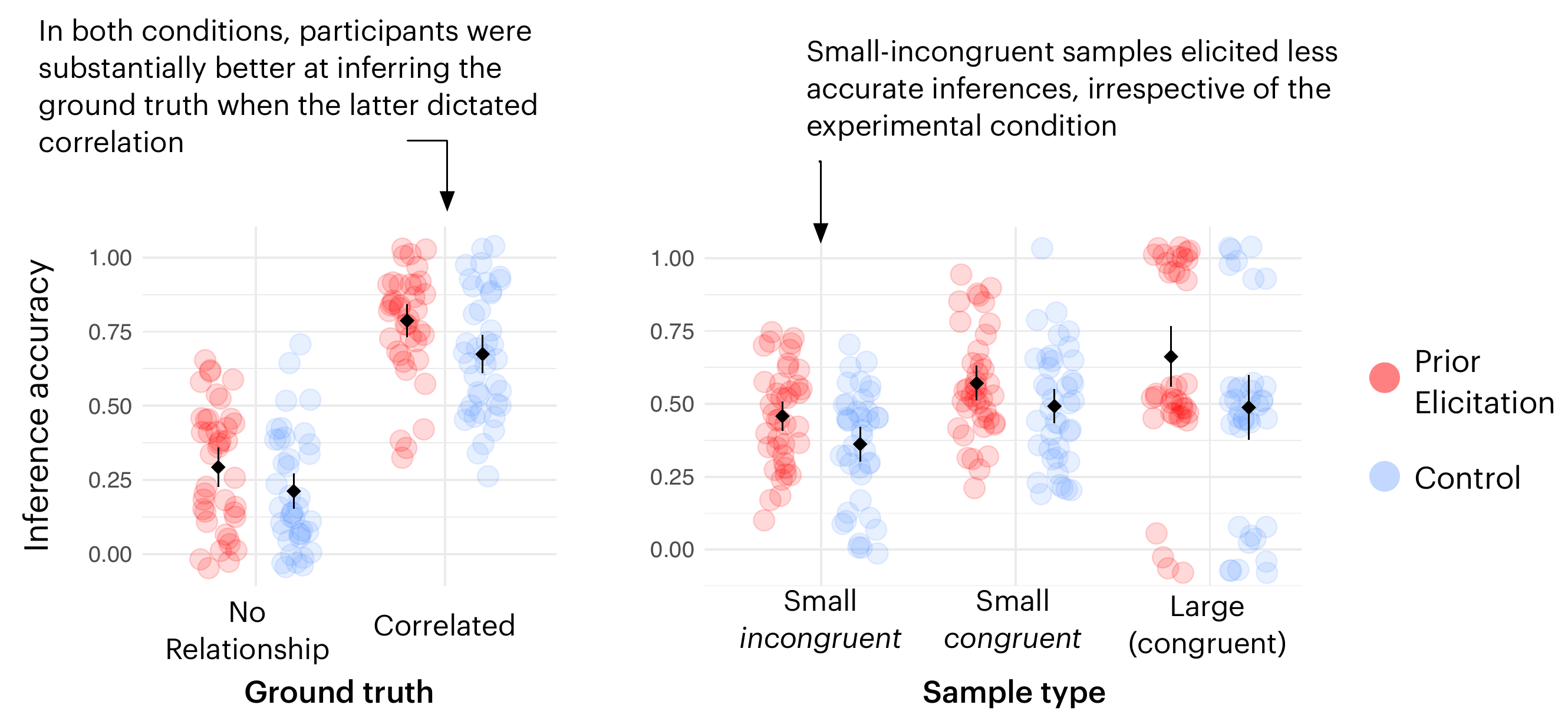}
  \caption{
  Inference accuracy by ground truth (\textbf{left}) and sample type (\textbf{right}). Dots (jittered to reduce overlap) depict mean accuracy for individual participants. Diamonds represent group means ($\pm95\%$ CI). Annotations are a summary of the significant effects. }
  \label{fig:exp1_sample_truth}
  \Description{
  Inference accuracy by ground truth and sample type. The left figure shows a vertical scatterplot comparing inference accuracy (on the y-axis ranging from 0.00 to 1.00) between the Prior Elicitation and Control groups for no-relationship and correlated ground truths. Dots represent the accuracy for individual participants and diamonds depict mean condition accuracy (with 95\% confidence intervals). The right figure shows a vertical scatterplot comparing inference accuracy (on the y-axis ranging from 0.00 to 1.00) between the Prior Elicitation and Control groups for small-incongruent, small-congruent, and large samples respectively. Dots represent the accuracy for individual participants and diamonds depict mean condition accuracy (with 95\% confidence intervals). Both figures are also annotated with a key finding at the top. The left figure is annotated "In both conditions, participants were substantially better at inferring the ground truth when the latter dictated correlation". The right figure is annotated "Small-incongruent samples elicited less accurate inferences, irrespective of the experimental condition".
  }
\end{figure*}

\begin{figure*}[t]
\center
\includegraphics[width=1\linewidth]{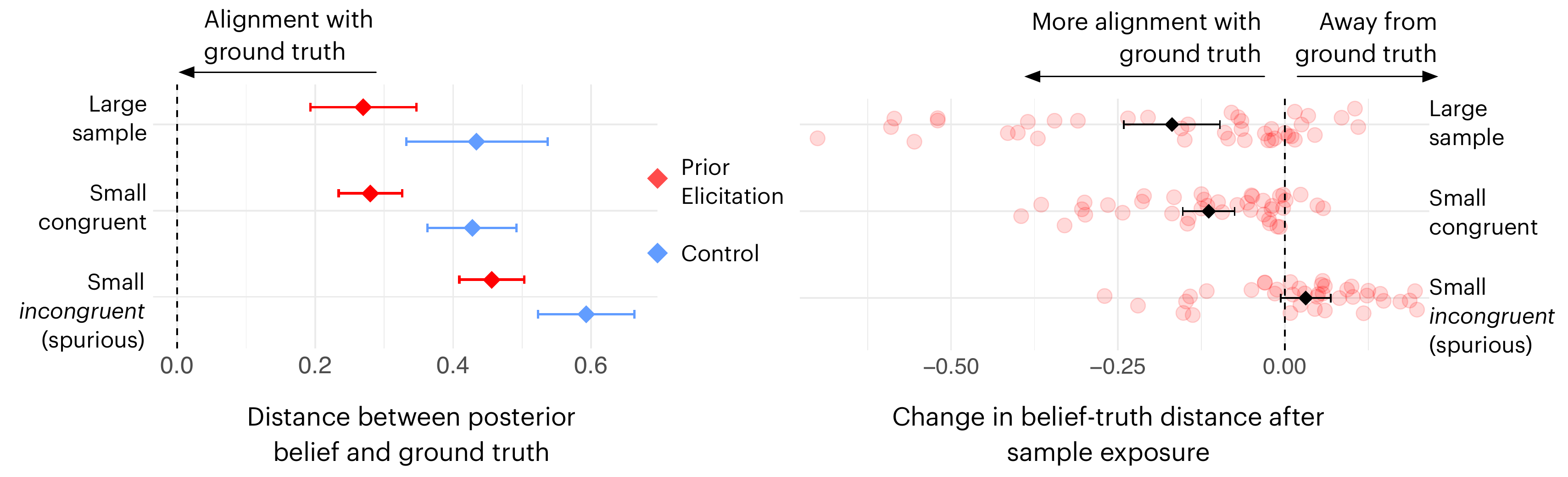}
  \caption{
  \textbf{Left: }Mean unsigned distance between the data-generating model ($\mu$) and elicited posterior slope. \textbf{Right: } Change in belief-ground truth distance from the prior to the posterior. Error bars represent 95\% CIs. Note that the right chart is limited to participants in the intervention (priors were not elicited in Control by design).}
  \label{fig:exp1_belief}
  \Description{
  Post-hoc analysis for Experiment 1. The left figure shows a distribution chart comparing the distance between posterior belief and ground truth (on the x-axis) for each user group across three sample types on the y-axis (Large, Small, and Small incongruent). The diamond represents the mean value for the group and the 95\% confidence interval is also included. There is an annotation at the top saying "Alignment with ground truth" with an arrow pointing to the left. The right figure shows a distribution chart with individual data points comparing changes in the belief-truth distance after data exposure (on the x-axis) across sample types on the y-axis (Large, Small, and Small incongruent) for the Prior Elicitation group. The diamond represents the mean value for the group and the 95\% confidence interval is also included. There are two annotations at the top. The first annotation says "More alignment with ground truth" with an array pointing to the left. The second annotation says "Away from ground truth" with an arrow pointing to the right.
  }
\end{figure*}

We found a significant main effect of sample type ($\chi^2(4)=20.39, p<.001$). Participants had better odds of inferring the correct model when viewing a small-congruent versus an incongruent sample (odds ratio: 1.99, CI: 1.45---2.73, $Z=4.28, p<.001$). Other differences between large versus small samples were not significant. Figure~\ref{fig:exp1_sample_truth}-right illustrates this relationship. We did not find evidence of interaction between the experimental condition and sample type ($\chi^2(2)=1.82, p=.4$).  Participants in the Prior Elicitation group were consistently better than Control at inferring the true relationship, regardless of sample size or its congruence with the data-generating model.

Lastly, we looked for differences in how well participants responded to questions across the two ground-truth types (i.e., correlation vs. no relationship). We found a significant main effect of ground truth  ($\chi^2(2)=21.7, p<.001$). Relative to a no-relationship baseline, trials with correlated variables were far more likely to elicit a correct inference (odds ratio: 11.47, CI: 5.58---23.6, $Z=6.62, p<.001$). Figure~\ref{fig:exp1_sample_truth}-left illustrates this effect. There was no interaction between the experimental condition and model type ($\chi^2(1)=0.13, p=.72$). Participants in the Prior Elicitation group were consistently better at recovering the ground truth regardless of whether the variables were correlated or not, even though the odds were much higher with true correlations.

\begin{figure*}[t]
\center
\includegraphics[width=.825\linewidth]{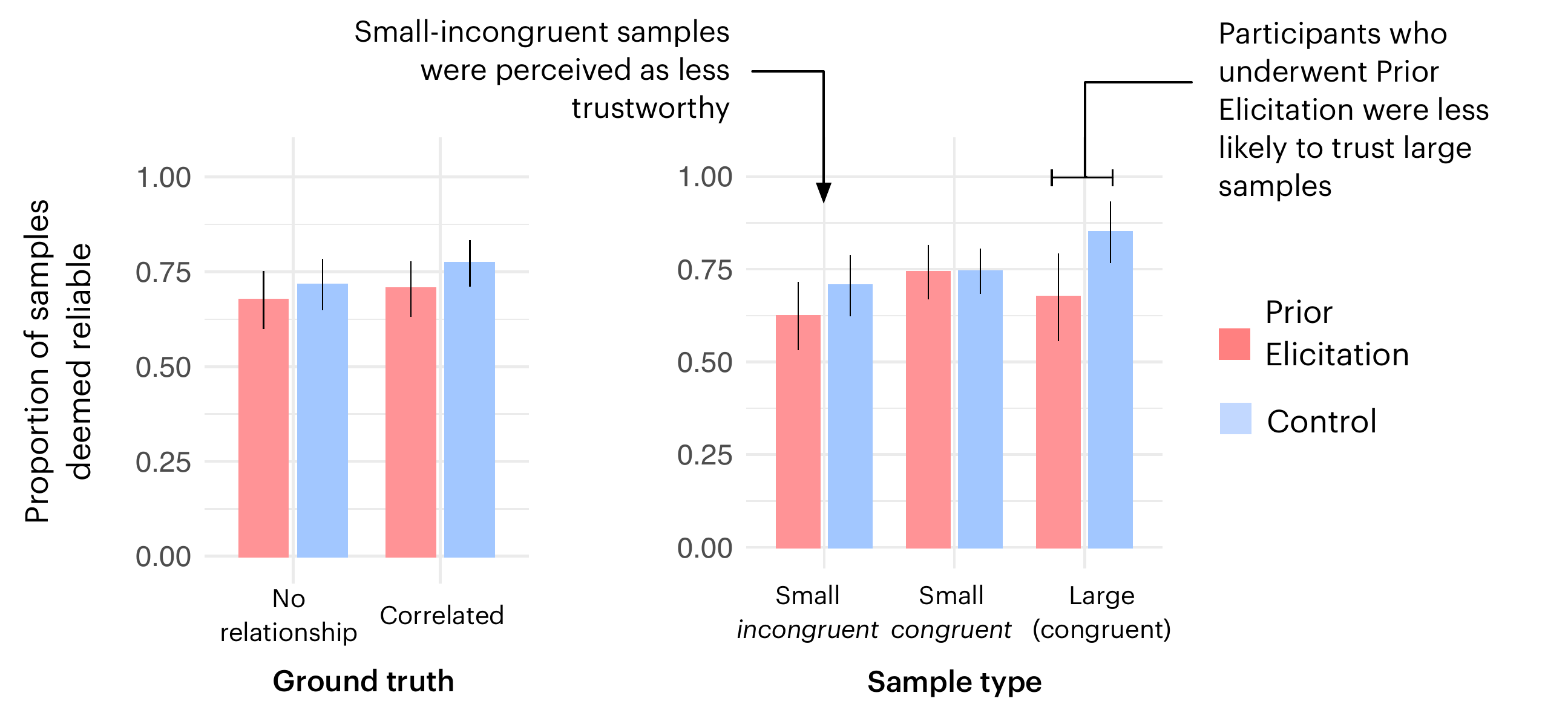}
  \caption{
  The proportion of samples deemed reliable by participants. Error bars are 95\% confidence intervals. }
  \label{fig:exp1_trust}
  \Description{
  The proportion of samples deemed reliable by participants. The left figure shows a paired bar chart comparing the proportion of samples deemed reliable (on the y-axis) by participants between the Prior Elicitation and Control groups for no-relationship and correlated ground truths (on the x-axis). The right figure shows a paired bar chart comparing the proportion of samples deemed reliable (on the y-axis) by participants between the Prior Elicitation and Control groups for small-incongruent, small-congruent, and large samples respectively. The 95\% confidence interval is included at each bar. The right figure is also annotated with two key findings at the top. The first  annotation points toward the first pair of bar charts with the description "Small-incongruent samples were perceived as less trustworthy". The second annotation points toward the third pair of bar charts with the description "Participants who underwent Prior Elicitation were less likely to trust large samples".
  }
\end{figure*}

\subsubsection{\underline{Posterior Beliefs and Belief Update (post-hoc analysis)}}

In addition to our primary metric of inference accuracy, we also measured the alignment between participants' posterior beliefs and the ground truth. This was computed by taking the absolute difference between the slope of the data-generating model ($\mu$) and the posterior slope (recall that the latter was elicited graphically in both conditions). Figure~\ref{fig:exp1_belief}-left compares the mean posterior distance for the two conditions under different sample types. Participants who underwent prior elicitation arrived at posteriors that aligned more closely with the ground truth ($\chi^2(3)=15.47, p<.01$). Sample type also had a significant main effect ($\chi^2(4)=97.23, p<.001$), although there was no evidence of an interaction with the experimental condition ($\chi^2(2)=0.25, p=.885$), suggesting consistently better belief alignment for the intervention group. Figure~\ref{fig:exp1_belief}-right further illustrates this effect by plotting the \emph{change} in belief-truth distance from the prior to the posterior (intervention group only). Although many participants appear to still have been misled by spurious samples, prior elicitation appears to induce better posterior approximation across all sample configurations.

\subsubsection{\underline{Perceived Sample Reliability}}

We modeled the likelihood of participants trusting a sample. Recall that, for each visualized sample, participants were asked whether they thought the sample was reliable. The main effect of the experimental condition was not significant ($\chi^2(4)=8.39, p=.08$). Participants in both conditions exhibited comparable levels of trust towards the samples shown to them. There was, however, a significant main effect of sample type ($\chi^2(4) = 18.83, p<.001$). Participants deemed large samples more trustworthy than small, incongruent samples (odds ratio: 1.99, 95\% CI: 1.23---3.21, $Z=2.82, p<.05$). Similarly, participants were also more likely to trust the small-congruent over the incongruent samples (odds ratio: 1.56, CI: 1.17---2.08, $Z=3.03, p<.01$). We also found a significant interaction between the experimental condition and sample type ($\chi^2(2)=6.82,p<.05$). Compared to their counterparts in Control, participants in the Prior Elicitation group were significantly more skeptical of large samples (odds ratio for trust: 0.31, CI: 0.12---0.82, $Z=-2.36, p<.05$). There was no evidence for the main effect of ground truth ($\chi^2(2)=2.78, p=.25$), or for its interaction with the experimental condition ($\chi^2(1)=0.39, p=.53$).

\subsection{Discussion}

The results show that eliciting analyst beliefs prior to them observing visualizations can improve inference. Participants in the Prior Elicitation group were approximately 21\% better than Control at recovering the ground truth. Alongside the improved inference, we observed a moderate reduction (approximately 12\%) in the rate of false discovery. Quantitative analysis of posterior beliefs reinforces this finding: those who underwent the intervention ended up with beliefs that are significantly closer to the ground truth (see Figure~\ref{fig:exp1_belief}-left). These results support H1.

As expected, participants made more errors when observing spurious (i.e., truth-incongruent) visualizations. They also made more false inferences when there was no  relationship between the variables, reflecting a potential bias to try to `discover' something in the data, rather than reporting a null result. Yet, these effects manifested similarly in the two experimental conditions. Thus, while prior elicitation broadly improves inference across all sample types, there is no evidence that the intervention is especially useful as an implement against spurious visualizations. Instead, we found better accuracy across all levels of sample configuration and ground-truth models.

Lastly, comparing participants' perceptions of sample reliability, there is weak, non-significant evidence  ($p=.08$) that prior belief elicitation makes people more skeptical. Counterintuitively, this effect seems more pronounced, presenting as significant with large samples (see Figure~\ref{fig:exp1_trust}-right). The results provide weak support to H3, although the increased skepticism towards the seemingly more robust (i.e., large-sample) visualizations is unexpected. It is possible that participants whose priors were elicited suffered anchoring effects, making them less trusting of even large samples when those turned unexpected evidence. Anecdotally, we came across responses that would suggest an anchoring effect in the Prior Elicitation group, leading participants to disregard robust evidence. For example, P-2707 responded to seeing a large sample with ``No, the sample didn't match with my belief. And I think that the sample is wrong in some sense.'' That said, this experiment was deliberately underpowered in the number of large samples it employed (under the assumption that smaller samples would pose a higher risk of false discovery). In a follow-up experiment, we employ a design with equal numbers of large- and small-sample stimuli. We also test interventions that could reduce confirmation bias.

\section{Experiment II}

\begin{figure}[t]
\center
\includegraphics[width=1\linewidth]{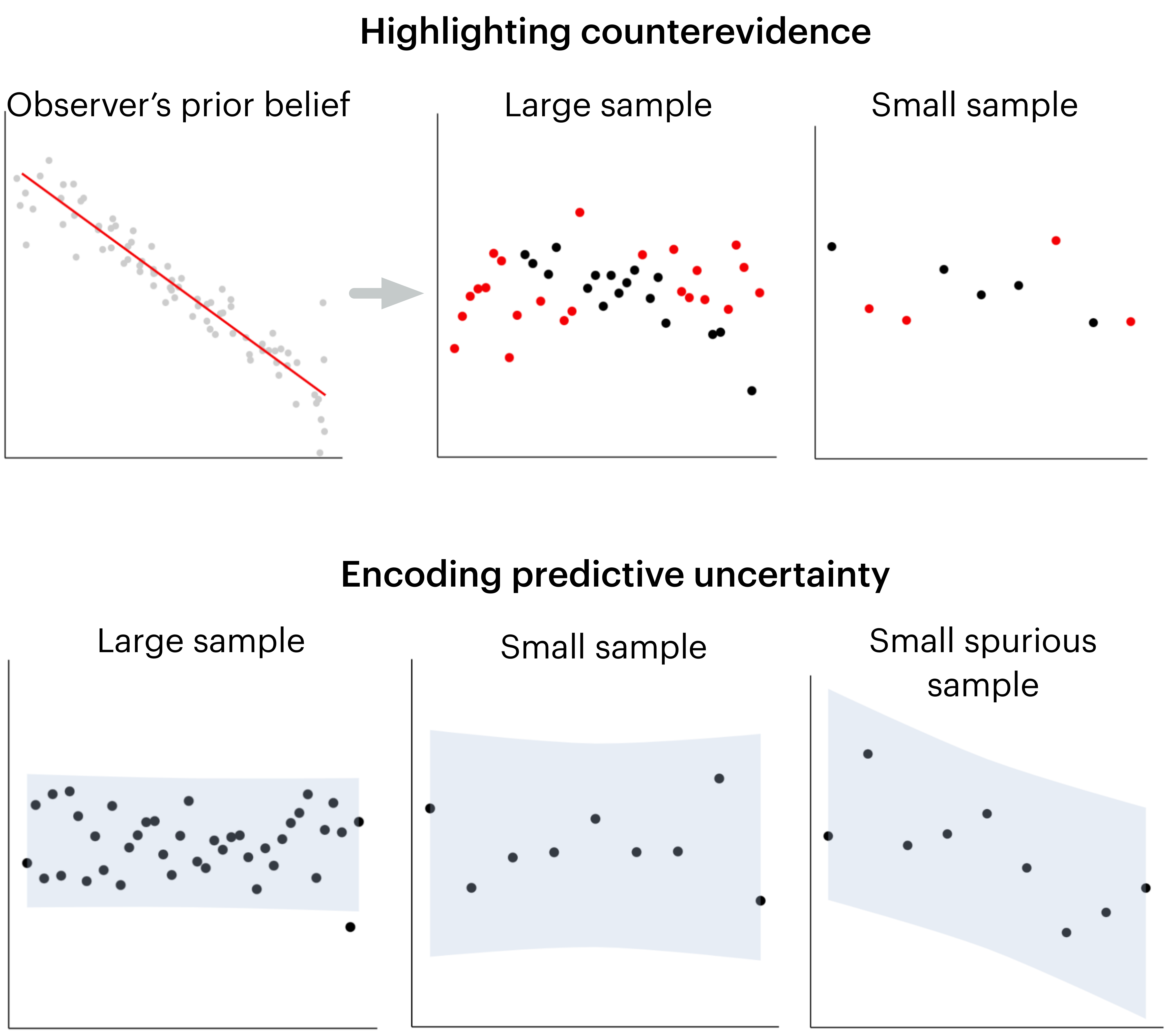}
  \caption{
    Two additional interventions tested in Exp.~2: highlighting sample observations  that deviate substantially from the observer's prior (top), and encoding the sample's predictive uncertainty, defined as the 95\% prediction interval of a linear fit to the sample.}
  \label{fig:exp2_interventions}
  \Description{
  Experiment 2 interventions. The top row illustrates the intervention for ``highlighting counterevidence'', with a participant's prior belief on the left, a large sample in the middle, and a small sample on the right. The left chart shows a scatter plot with a negative slope as the participant's prior belief. For the middle and right charts, data points that deviate from the belief are highlighted in red while the rest are black. The bottom row illustrates the ``predictive uncertainty'' intervention, with a large sample on the left, a small sample in the middle, and a small spurious sample on the right. The left and middle charts show data points with no relationship as well as their corresponding predictive uncertainty areas. The right chart shows data points with a negative relationship and their corresponding predictive uncertainty areas. 
  }
\end{figure}

Experiment 1 shows that externalizing priors could improve visual inference. But the results also raised concerns about a potential anchoring effect or confirmation bias. Specifically, visualizing beliefs side-by-side with samples may provide an `excuse' for participants to reject incompatible data, or at least overweight their beliefs. In this experiment, we test two additional interventions to reduce the chance of such bias, by alerting participants to evidence that is incompatible with their beliefs. This is achieved by perceptually highlighting observations in a sample that are inconsistent with viewer expectation. Specifically, we employ a salient color (red) to emphasize points in a scatterplot that deviate significantly from the expected linear model specified by the observer. An observation is considered a significant deviation if it is located more than $3\sigma$ away from the elicited trendline. Moreover, we  consider another alternative intervention that features the uncertainty implied by the sample. Specifically, we fit the sample to a linear model and visualize the 95\% prediction intervals of that model as shaded uncertainty bounds. Figure~\ref{fig:exp2_interventions} illustrates the two interventions.

The two interventions could aid observers in deciding how much to weigh their prior beliefs against potentially noisy samples. The \textbf{Highlight} condition cues the observer onto the \emph{amount} of counterevidence. A large sample that is incompatible with one's expectation is likely to pop out and invite more attention than a smaller one. The conspicuous highlight could also make it difficult for the observer to ignore unexpected evidence, especially if large. On the other hand, the \textbf{Uncertainty} condition encodes the sample's usefulness in \emph{predicting future observations}. Larger samples would exhibit smaller uncertainty bounds, which participants may interpret as an indicator of sample robustness. In both of these interventions, we elicit participant beliefs \emph{prior to} and \emph{after} exposing them to a sample (i.e., similar to the Prior Elicitation condition in Exp.~1). In addition to the two above interventions, we also include a third \textbf{Baseline} condition that is identical to the intervention in Exp.~1.

\subsection{Hypotheses}

We expected improved inference with the two new interventions. We also expected an increase in participants' trust in the large samples:

\textbf{H4} --- Participants will benefit from Highlighting evidence inconsistent with their beliefs. Similarly, we expect benefits from explicitly visualizing a sample's predictive Uncertainty. We anticipate both of these interventions to lead to better inference compared to participants in the baseline condition. 

\textbf{H5} --- Compared to Baseline, participants observing samples with Highlights or with Uncertainty annotations will be more trusting of large samples.

\subsection{Participants, Experimental Design, and Procedures}

We recruited 120 participants (48 males, 72 females; mean age of 34.8 years) from Amazon Mechanical Turk. We recruited workers who are US residents with a minimum task-approval rate of 98\%. Participants were compensated with a \$5 payment. We used a between-subject design, with an equal number of participants (40) randomly assigned to each of the three conditions (Highlight, Uncertainty, or Baseline). The experiment consisted of 16 trials. In each trial, participants responded to a prompt question and followed the same steps as in Exp.~1: first visually providing their prior belief about the relationship between two variables, then observing a scatterplot sample, and lastly providing their open-ended inference and posterior belief (see Figure~\ref{fig:stimulus} for illustration of these steps). All participants in the experiment had to externalize their beliefs before and after observing a sample. For the two new interventions, the sample showed a highlight of data points deviating from the specified prior or displayed uncertainty bounds depicting predictive intervals from a linear fit of the sample. 

Unlike Exp.~1, we used an equal number of small and large samples in this experiment. Eight trials displayed large samples ($n=40$ data points each) whereas eight showed small samples ($n=9$). Of the 8 small-sample trials, 3 were incongruent with the ground truth, for a spurious sample rate of 37.5\% (consistent with the model likelihood; see Table~\ref{tab:simulation}). Trial order was randomized. Additionally, we randomly assigned the large, small-congruent, and small-incongruent samples to the 16 unique questions. However, we balanced the incongruent samples such that each participant would encounter one spurious visualization with each of the three model types (positive, negative, no correlation).

\begin{figure}[t]
\center
\includegraphics[width=1\linewidth]{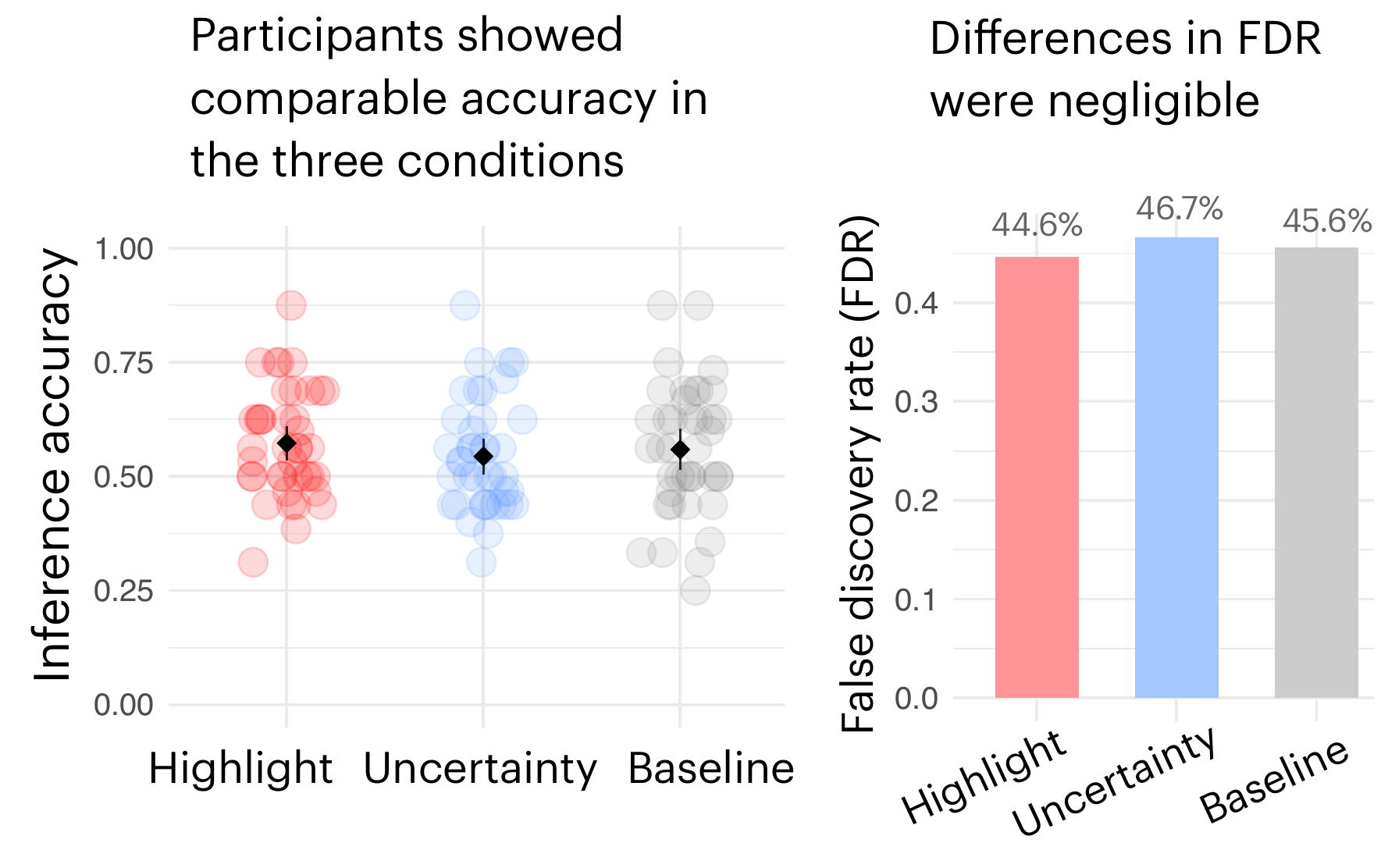}
  \caption{
  Inference accuracy in Exp.~2 by intervention (\textbf{left}) along with the corresponding false discovery rate.} 
  \label{fig:exp2_accuracy}
  \Description{
  Accuracy and false discovery rate for experiment 2: The left figure shows a vertical scatterplot comparing the inference accuracy (on the y-axis ranging from 0.00 to 1.00) for the Highlight, Uncertainty, and Baseline groups. Dots represent the accuracy for individual participants and diamonds depict mean condition accuracy (with 95\% confidence intervals). At the top of the figure, there is an annotation saying "Participants showed comparable accuracy in the three conditions". The right figure shows a bar chart comparing the false discovery rate (on the y-axis ranging from 0.00 to 1.00) for the Highlight, Uncertainty, and Baseline groups. Each of the three bar charts is annotated with the false discovery percentage corresponding to values on the y-axis (44.6\%, 46.7\%, and 45.6\% respectively). At the top of the figure, there is an annotation saying "Differences in FDR were negligible". 
  }
\end{figure}

\begin{figure*}[t]
\center
\includegraphics[width=1\linewidth]{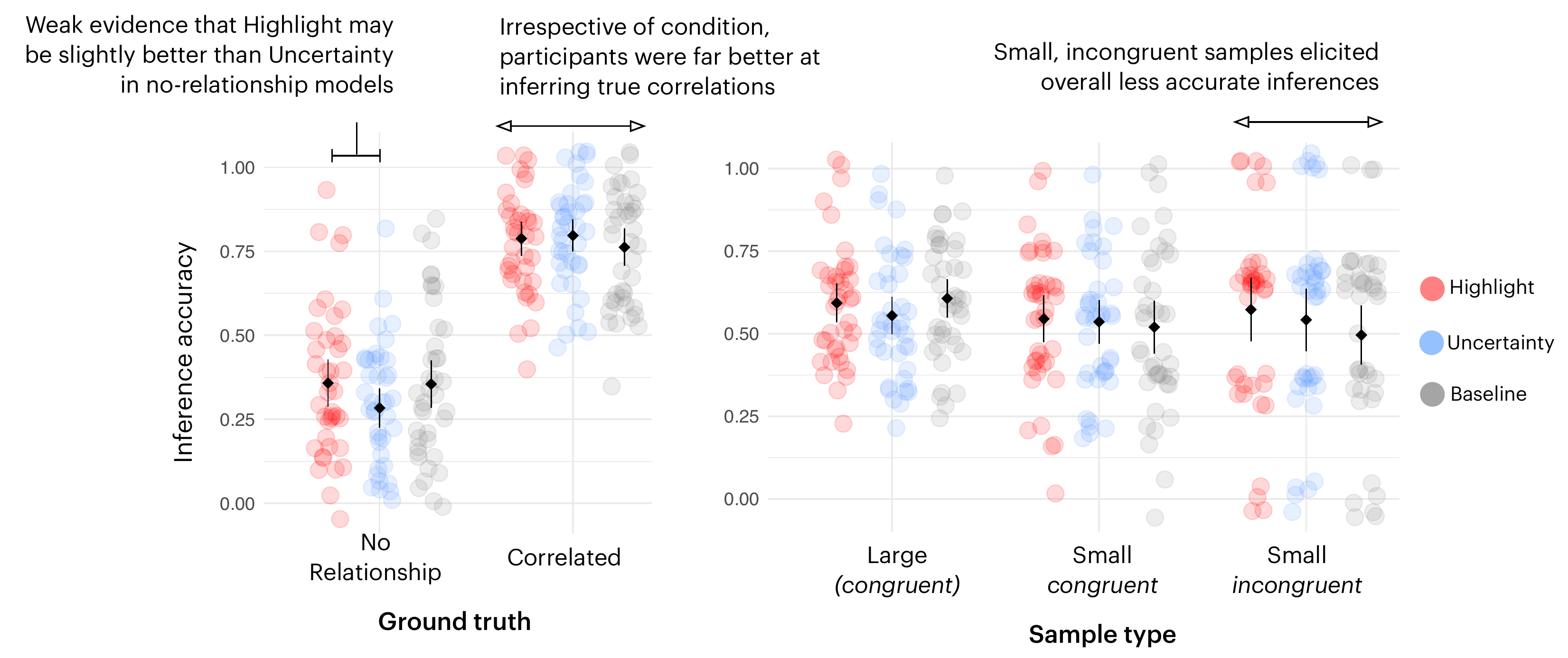}
  \caption{Inference accuracy in Exp.~2 by ground truth (\textbf{left}) and sample type. Points (representing rates for individual participants) were jittered slightly to reduce overlap.} 
  \label{fig:exp2_sample_truth}
  \Description{
  Inference accuracy by ground truth and sample type. The left figure shows a vertical scatterplot comparing inference accuracy (on the y-axis ranging from 0.00 to 1.00) between the Highlight, Uncertainty, and Baseline groups for no-relationship and correlated ground truths (on the x-axis). Dots represent the accuracy for individual participants and diamonds depict mean condition accuracy (with 95\% confidence intervals). At the top, there are two annotations for key findings. The first annotation points toward the no-relationship group with a description saying "Weak evidence that Highlight may be slightly better than Uncertainty in no-relationship models". The second annotation is above the correlated group saying "Irrespective of condition, participants were far better at inferring true correlations". The right figure shows a vertical scatterplot comparing inference accuracy (on the y-axis ranging from 0.00 to 1.00) between the Highlight, Uncertainty, and Baseline groups for small-incongruent, small-congruent, and large samples (on the x-axis) respectively. Dots represent the accuracy for individual participants and diamonds depict mean condition accuracy (with 95\% confidence intervals). There is an annotation above the Small incongruent group saying "Small, incongruent samples elicited overall less accurate inferences".  
  }
\end{figure*}

\subsection{Results}

Participants completed the experiment in 58.7 minutes on average. They provided 1,920 responses in total. We coded inference accuracy (correct or incorrect) using the same coding procedure as in Exp.~1 (see \S\ref{sec:coding}). We removed 34 responses (1.8\%) which we were unable to code, leaving 1,886 responses in the analysis.

\subsubsection{\underline{Inference Accuracy}}

 We fit the results to a logistic regression model to predict the correctness of  inferences. The model includes three main factors: the experimental condition (Highlight, Uncertainty, or Baseline), the sample type (large, small-congruent, and small-incongruent), and the ground-truth type (no relationship or correlation between the variables). We also included interaction terms to model variations in the latter two  due to different experimental conditions. Random intercepts for individual participants and questions were also included in the model, consistent with Exp.~1. Figure~\ref{fig:exp2_accuracy} illustrates the overall accuracy and false discovery rate for the three interventions. We did not find a significant main effect of the experimental condition ($\chi^2(8)=10.56, p=.23$). The three conditions elicited comparable inferences in terms of accuracy. The false discovery rate was also consistent (44.6\% to 46.7\%) and comparable to that seen in Exp.~1 (47.2\%).

We found a significant main effect of ground-truth type ($\chi^2(3) = 22.57, p < .001$). Trials with an underlying correlation were judged far more accurately compared to trials where the ground truth specified no relationship between variables (odds ratio: 11.7, CI: 4.9---27.9, $Z = 5.55, p < .001$). Figure~\ref{fig:exp2_sample_truth}-left illustrates this effect. The evidence of interaction between the experimental condition and ground truth was weak ($\chi^2(2)=5.08, p=.08$). The Highlight intervention \emph{may be} slightly better at preventing false discovery (odds ratio: 1.6, CI: 1.01---2.53) than the Uncertainty display when there is no true correlation, although this effect is not robust ($Z=2.01,p=.11$).

We found a significant main effect of sample type ($\chi^2(6)=32.445, p<.001$). Small, incongruent samples were less likely to elicit correct inferences compared to both large (odds ratio: 0.46, CI: 0.34---0.62, $Z = -4.99, p < .001$) and small-congruent samples (0.44, CI: 0.32---0.62, $Z = 4.75, p < .001$).  This effect is illustrated in Figure~\ref{fig:exp2_sample_truth}-right. Large and small-congruent samples were judged with comparable accuracy ($Z=0.26,p=.96$). We did not find evidence of an interaction between the experimental condition and sample type ($\chi^2(4)=3.93, p=.42$). Participants appear to perform similarly with all three interventions (Highlight, Uncertainty, and Baseline). Additional post-hoc analysis of posterior and prior beliefs can be found in the supplementary materials.

\begin{figure*}[t]
\center
\includegraphics[width=.8\linewidth]{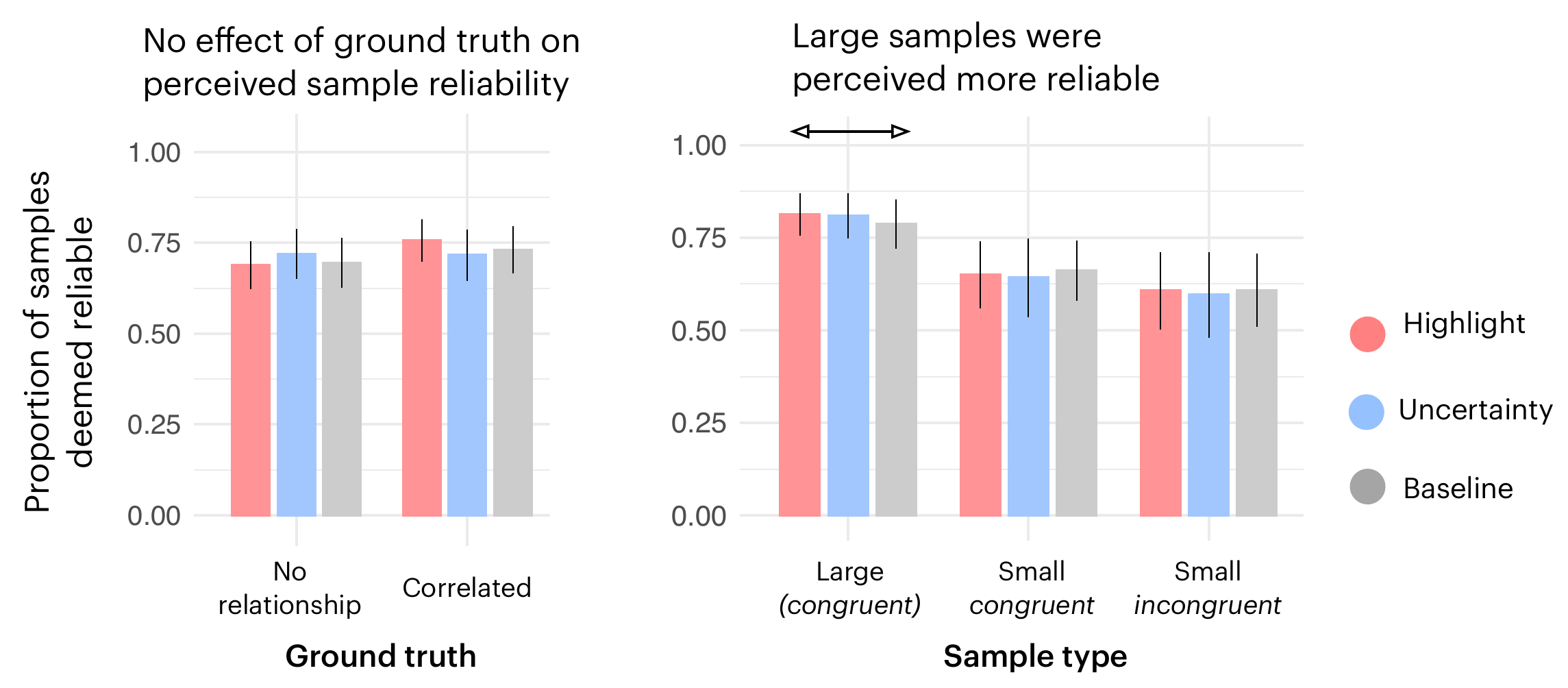}
  \caption{Mean perceived sample reliability in Exp.~2 by ground truth (\textbf{left}) and sample type. Error bars are 95\% CIs.} 
  \label{fig:exp2_trust}
  \Description{
  Mean perceived sample reliability by participants. The left figure shows a group bar chart comparing the proportion of samples deemed reliable (on the y-axis ranging from 0.00 to 1.00) by participants between the Highlight, Uncertainty, and Baseline groups for no-relationship and correlated ground truths (on the x-axis). There is an annotation at the top saying "No effect of ground truth on perceived sample reliability". The right figure shows a bar chart comparing the proportion of samples deemed reliable (on the y-axis ranging from 0.00 to 1.00) by participants between the Highlight, Uncertainty, and Baseline groups for small-incongruent, small-congruent, and large samples (on the x-axis) respectively. There is an annotation above the Large group saying "Large samples were perceived more reliable". The 95\% confidence interval is included at each bar. 
  }
\end{figure*}

\subsubsection{\underline{Perceived Sample Reliability}}
We analyzed participants' perceptions of sample reliability (a binary response). Figure~\ref{fig:exp2_trust} illustrates the proportion of samples deemed trustworthy. We did not find a significant main effect of the experimental condition ($\chi^2(8)=2.91, p=.94$). There was a significant main effect of sample type ($\chi^2(6) = 82.93, p<.001$), with participants trusting large samples more than the small-incongruent (odds ratio: 3.22, CI: 2.41---4.29, $Z = 7.92 , p < .001$) and small-congruent samples (odds ratio: 2.39, CI: 1.87---3.07, $Z = 6.87, p < .001$). But there was no evidence for this effect varied at all between the three conditions  ($\chi^2(4)=1.04,p=.9$). We found no evidence of an effect for ground-truth type ($\chi^2(3)=4.59, p=.21$) or for its interaction with the intervention ($\chi^2(4)=1.04,p=.9$). 

\subsection{Discussion}

Overall, we did not see the anticipated effects for highlighting counterevidence (i.e., observations inconsistent with the observer's model), or for explicitly encoding the sample's predictive uncertainty. Inference accuracy across the three conditions was quite comparable. At best, any difference between the interventions is likely small and potentially limited to a specific context. An example where such an effect might have presented is when the ground truth dictated no relationship. Here, highlighting counter observations could be slightly better at preventing false discovery than displaying an uncertainty representation (see Figure~\ref{fig:exp2_sample_truth}-left). This could be due to a difficulty by participants in reading the uncertainty intervals which are often misinterpreted~\cite{belia2005researchers}. By comparison, it may be easier to comprehend a sample's implication in the Highlight condition: an unexpected scatterplot presenting a larger number of counter-observations should invite more scrutiny, possibly causing the observer to be more apprehensive. That said, the effect is likely small (odds ratio: 1.01---2.53) and  not statistically robust. The results do not provide support for H4. In terms of perceived sample reliability, the rates were quite comparable in the three conditions. As expected, participants saw the large samples as significantly more trustworthy -- an effect that held across all interventions. The lack of significant interaction suggests that perceived sample reliability is unaffected by any of the additional interventions.  The results thus do not provide support for H5 either. 

The null results in this experiment could be because the interventions are, after all, largely similar. All three experimental conditions included prior and posterior elicitation, with the main difference being in how the sample was presented. It would appear that the latter has a marginal impact when weighed against the primary interaction (i.e., belief elicitation). Despite the lack of a robust effect, the experiment replicated much of the patterns observed in Exp.~1. For example, participants in both experiments were more likely to  infer the ground truth when the latter dictated correlation. Small, incongruent samples also elicited less accurate inferences in both experiments. Notably, this experiment included an equal number of small and large samples, with the latter perceived as more trustworthy by participants. The experiment thus lends additional validation to the findings of Exp.~1 and to the general methodology.

\section{General Discussion}

Exploratory data analysis (EDA) enables analysts to characterize useful patterns in a way that is not strictly limited by a formal model. However, in addition to surfacing true patterns, EDA can amplify noise. Analysts can sometimes overinterpret these spurious signals, causing them to see relationships that are not truly there, or that which might not generalize beyond the sample at hand. The flexibility afforded by modern visualization tools exacerbates this problem by encouraging a large number of visual tests or model checks~\cite{zgraggen2018investigating,pu2018garden}. While multiple comparisons are often dealt with in formal statistics, few interventions have been proposed to control the false discovery rate in interactive analyses.

\subsection{Belief Elicitation Leads to Better Inference from Visualized Data}

One approach to guard against spurious signals is to activate viewers' prior knowledge and nudge them to be critical of apparent patterns they encounter in visualizations. Earlier work shows that belief-driven interactions can indeed promote good visual analytic practices~\cite{koonchanok2021data}. Our findings suggest that these affordances also help people infer better models about the true state of the world. Participants whose beliefs were elicited prior to data exposure articulated 21\% more accurate inferences. They also made 12\% fewer false discoveries. Interestingly, these benefits were not limited to responses made with noisy visualizations. Rather, participants who furnished their priors were consistently better at inferring the real data-generating process, from both reliable and spurious samples.

We suspect two mechanisms behind the improved inference. First, activating one's prior knowledge can encourage them to be more skeptical, particularly in the face of improbable data. Exp.~1 provided weak evidence of increased skepticism among those whose priors were elicited. Though many were still misled by incongruent samples, participants in the intervention group seemed more adept at tempering the influence of spurious visualizations, ultimately arriving at more accurate posteriors. A second possible explanation is that, by prompting observers to reflect on what they know, we trigger deeper processing of data. Kim et al. suggest similarities between belief articulation and self-explanation, wherein a learner generates an explanation to themself as they attempt to make sense of new information~\cite{kim2017explaining}. Self-explanation often leads to better learning~\cite{vanlehn1992model,rittle2017eliciting} even when induced externally (e.g., by a teacher)~\cite{bisra2018inducing}. Similar cognitive effects may unfold when observers are prompted to record their beliefs. The enhanced processing could improve one's understanding of sample implications, which may in turn translate to better inference. That the advantage was observed in all sample configurations supports this idea.  We speculate both of the above-mentioned factors  (data skepticism and enhanced processing) come into play when analysts operationalize their prior knowledge during graphical inference.
	
\subsection{Confirmation Bias and the `Law of Small Numbers'}

Belief elicitation appears to help reduce the chance of false discovery, although the intervention could theoretically lead to an anchoring effect. We saw potential signs of this in Exp.~1. Specifically, participants whose priors were elicited seemed less trusting of large samples. Moreover, even with the benefit of the intervention, large samples induced correct inference only 66\% of the time --- a lower rate than what might be expected from a normative analyst. This phenomenon may indicate a non-belief in the law of large numbers; people often underestimate the evidential significance of a large sample while overstating the importance of a small one~\cite{benjamin2016model,tversky1971belief}. It may also reflect confirmation bias, where an analyst persists with their belief despite strong evidence to the contrary. We investigated two additional interventions in Exp.~2, aiming to alleviate the above issue. We expected these interventions to increase the likelihood of correct inference, particularly from large samples. Instead, we found no reliable effects for  highlighting counterevidence, nor for encoding a sample's predictive uncertainty. The results suggest that the benefits of belief elicitation are decoupled from the manner in which the sample is represented, as long as the observer can still see the raw sample. 

\subsection{Implications for Visualization Tools}

\hl{Our work suggests benefits for operationalizing analyst beliefs during visual analysis. To our surprise, the advantage held across a variety of sample characteristics (i.e., large, small, and even spurious samples). These results have important implications for the design of visualization tools. We consider the feasibility of broadly implementing this intervention and the potential impact it might generate. We also reflect on future interaction designs for fostering trustworthy visual analytics.}

\subsubsection{\underline{Intervention Feasibility and Potential Impact}}

\hl{While it might take additional effort to externalize one's belief, our findings suggest that the overhead is small. Participants in the intervention and control conditions took similar times, indicating that graphical elicitation requires minimal added effort. Yet, the intervention yielded a tangible improvement in inference. Consequently, we might envision the intervention as a standard feature in visualization tools. The minimally intrusive nature of the interaction also suggests that it would be embraced and utilized frequently by analysts. Although `required' in our setup, belief elicitation can also be implemented as an optional feature. For example, in a system developed by Koonchanok et al., analysts can choose to paint their prior expectations into charts if they wish or, alternatively, forgo this interaction altogether in favor of immediately seeing data, as is the case in current visualization systems.}~\cite{koonchanok2021data}. \hl{Depending on the chart, the system could support a variety of elicitation techniques (e.g., a paintbrush to specify expected pointcloud density in scatterplots or multi-variate ribbons for parallel coordinates). Based on our findings, we can anticipate the presence of these interactions to improve graphical inference.}

\hl{It is important to also investigate the potential side effects of such interventions. For example, eliciting beliefs could cause analysts to overly fixate on their existing hypotheses. This in turn may lead to unexpected patterns being missed. Research indeed suggests that while hypotheses provide useful constraints for sensemaking}~\cite{klahr1988dual,dunbar2001scientific}, \hl{they can also be a liability in that they limit one from considering unexpected data outside of their focus}~\cite{yanai2020hypothesis}. \hl{Similar dynamics might come into play with interactive visualizations: in one experiment, analysts who specified their hypotheses entertained fewer visualizations in their EDA process and ultimately reported fewer observations}~\cite{koonchanok2021data}. \hl{Of course, fewer `insights' might not be a bad thing, especially when such insights align more closely with the real data-generating processes.} It is important, however, to consider other biases analysts may be subject to, such as structural incentives to increase discovery~\cite{grimes2018modelling,kiai2019protect}. This could still motivate questionable behaviors like p-hacking despite tool-level interventions~\cite{head2015extent}. 

\subsubsection{\underline{Future Design Directions}}

\hl{Our work suggests future research avenues and design interventions for the visualization community to consider. One important goal is promoting balanced visual analytic workflows and practices. For example, tools can help analysts explicitly test their beliefs while also nudging them to explore patterns that are outside the purview of their existing hypotheses. This can be done by augmenting recommender algorithms to consider analyst hypotheses}~\cite{wongsuphasawat2015voyager}, \hl{thereby allowing systems to respond with relevant plots as well as with other interesting visualization users might not otherwise see given their foci. Visual analytics tools may also be able to better predict intents by accounting for both user priors and their interaction histories. With this information, systems can nudge analysts to adopt a broader exploratory stance when called for, or conversely, steer them to be more hypothesis-centric. The appropriate balance between openness and directedness will depend on the context as well as the potential risks (e.g., of false discovery). To that end, augmenting metrics of analytic breadth and depth}~\cite{feng2018patterns} \hl{to incorporate analyst knowledge may open up new ways for supporting varying analysis styles and needs.}

\hl{Visualization systems can also help users reason about data-generating processes with simulated data drawn either from null}~\cite{buja2009statistical} \hl{or prior models}~\cite{hullman2021designing}. \hl{While there are successful examples here (most prominently the lineup protocol}~\cite {buja1987data}), \hl{the  design space remains underexplored. For example, in addition to showing users entire simulated datasets, we might present summaries of those simulations, highlighting the portion of extreme visualizations given certain assumptions, or grouping simulation results by visual characteristics (e.g., via scagnostics}~\cite{wilkinson2005graph}). \hl{An analyst can inspect those summaries at a high level to understand the range of possible visualization outcomes, given an uncertain data-generating process. Similar ideas are already common in statistics (e.g., using bootstrapping and prior/posterior predictive simulation). However, the potential analogs for these methods in visualization tools are yet to be fully explored.}

\section{Limitations}

Although robust in design, there are limitations to our study that should be contextualized. First, our participants were recruited from Amazon Mechanical Turk. It is fair to assume that this sample does not represent professional data analysts. Evidence does suggest that even experienced analysts could benefit from belief-driven interactions in visualization tools~\cite{koonchanok2021data}. Still, given the crowdsourced nature of this study, the effects we found may not carry over, or at least present with similar magnitudes in seasoned visualization users, especially those who are trained in statistical inference. A second  limitation is the controlled nature of the experiments; stimuli were presented in a predetermined order, with participants having no say over what visualizations to inspect. By comparison, real EDA is fluid with analysts typically deciding what projections to look at on the fly based on emergent data features and evolving goals~\cite{reda2016modeling,battle2019characterizing}. Future work could attempt to replicate our findings in setups that allow for more naturalistic EDA. A~related issue stems from the ground-truth models we developed, the parameters of which were also crowdsourced. Therefore, while much of the samples observed by participants were spurious (and hence misleadingly surprising), the underlying truth was consistent with the crowd wisdom. The average participant may thus have had a reasonable chance of making correct inferences by simply falling back to their intuition. To be fair, this advantage was present in all  interventions tested. That said, it would be interesting to replicate our study using ground-truth models that are inconsistent with crowd wisdom. Lastly, we based our study on simplified linear models that dictate correlations between two variables. This particular prompt format and the corresponding belief elicitation device were inspired by earlier studies~\cite{karduni2020bayesian}. Yet, this format  presents a limited space of visual inferences one can make. We also used a  coding scheme that classifies inferences as either `correct' or `incorrect', in effect assuming a dichotomous interpretation of visualizations.  Future work could also consider more expansive criteria for gauging inference quality, beyond the binary approach we used in this work. It would also be interesting to replicate our experiments with other kinds of models, and with other chart types and belief elicitation devices.

\section{Conclusion}

In the process of exploratory visual analysis, people sometimes draw inferences that are intended to generalize. Such inferences may be based on spurious visual patterns as opposed to true underlying effects or relationships. We explored interventions to combat the problem of spurious discovery in visual analytics. Specifically, we investigated if eliciting and activating analyst working knowledge can improve inference quality. Participants whose beliefs were elicited before observing samples made significantly more accurate inferences and exhibited a lower false discovery rate. The effect was robust when observing both true and spurious plots, suggesting broad utility to the intervention. Additional interventions for visually highlighting counterevidence or sample uncertainty did not yield significant results over the baseline effect above. The findings suggest that visual inference can be improved with relatively lightweight interactions that can be easily incorporated into visualization systems. Such interventions could have concrete, if moderate, success in addressing the problem of false discovery.

\begin{acks}

We would like to thank Abhinav Sikharam, Vijay Mittal, and Amey Salvi for their help in developing the experiment interface. This paper is based upon research supported by the National Science Foundation under award 1942429. 

\end{acks}

%\section{Acknowledgments}

\interlinepenalty=10000
\balance
\bibliographystyle{ACM-Reference-Format}
\bibliography{00_bib}

\end{document}